\documentclass[a4paper,11pt]{article}

\usepackage[english]{babel}
\usepackage[utf8]{inputenc}
\usepackage{jcappub} % for details on the use of the package, please see the JINST-author-manual
\usepackage{amsmath}
\usepackage{empheq}
\usepackage{mathtools}
\usepackage{slashed}
\usepackage{graphicx}
\usepackage[]{hyperref}
\usepackage{xcolor}

\usepackage{bm}
\usepackage{epsfig}
\usepackage{bbold}
\usepackage{graphicx}
\usepackage{amsmath}
\usepackage{amssymb}
\usepackage{amsbsy}

\usepackage{subfigure}
\usepackage{nicefrac}
\usepackage{slashed}
\usepackage{afterpage}
\usepackage{psfrag}

\usepackage[capitalise]{cleveref}
\usepackage{dsfont}
\usepackage{comment}
\usepackage{hyperref}
\newcommand{\beq}{\begin{equation}}
\newcommand{\eeq}{\end{equation}}
\newcommand{\bea}{\begin{eqnarray}}
\newcommand{\eea}{\end{eqnarray}}

\newcommand{\niz}{n_{i}^{(0)}}
\newcommand{\DOz}{\Delta \Omega^{(0)}}

\title{The phenomenology of Axion Relic Pockets}
\author[a]{Wafaa Khater,}
\author[b]{M.C. David Marsh,}
\author[b]{Charalampos Nikolis}
\affiliation[a]{Department of Physics, Birzeit University, Palestine}
\affiliation[b]{The Oskar Klein Centre, Department of Physics,
Stockholm University, Stockholm 106 91, Sweden}

\emailAdd{wkhater@birzeit.edu}
\emailAdd{david.marsh@fysik.su.se}
\emailAdd{charalampos.nikolis@fysik.su.se}

\abstract{Axion relic pockets are phase-transition remnants consisting of regions of false vacuum stabilised from collapse by a hot axion gas. Axion relic pockets can comprise dark matter, but little is known about their phenomenology. We consider trapped axions coupled to electromagnetism  and derive the axion to photon conversion rate accounting for the pocket's compactness and spherical geometry. As a case study, we show that axion-photon conversion occurs in atomic electric fields, leading to high-energy electromagnetic cascades that may be detectable by neutrino, cosmic-ray, and dark-matter experiments. The distribution of electromagnetic showers is inherently isotropic, and upward-going showers provide a smoking-gun signal for the model. Using recent data from the IceCube neutrino telescope, we derive the first limits from terrestrial experiments on axion relic pockets as dark matter. However, for the simplest realisation of the model, only a small part of the parameter space can give signals strong enough to be constrained by current data. We also comment on the prospects for discovering axion relic pockets using fluorescence and radio Cherenkov detection by atmospheric instruments and balloon-borne payloads, as well as traditional dark matter direct detection experiments. These results 
provide the necessary tools to initiate phenomenological studies of axion relic pockets across different environments.
}

\begin{document}
\maketitle

\section{Introduction}
Despite the remarkable success of the Standard Model (SM) of particle physics, observations point to the need for a more complete framework that can address problems related to the nature of dark matter, dark energy and the quantisation of gravity.
Pseudo-scalar particles coupled to gauge bosons via a Chern-Simons term (axion-like particles),
and scalar fields (`dilatons' or moduli)  controlling the values of coupling constants, are commonly predicted in well-motivated SM extensions and are ubiquitous in low-energy effective field theories descending from string compactifications \cite{Svrcek:2006yi,Arvanitaki:2009fg}. The interactions of these particles with matter and their couplings are determined by the details of the UV theory they descend from (see e.g. \cite{Choi:2020rgn,Arza:2026rsl} for reviews).

\begin{figure*}[b]
    \centering
    \includegraphics[width=0.9\textwidth]{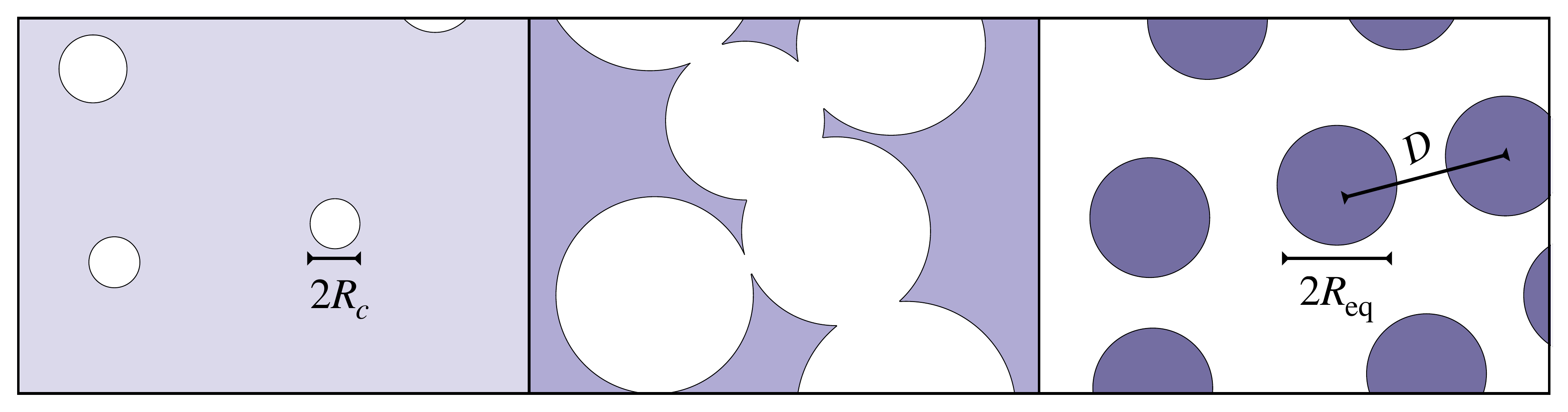}
\caption{Left: Nucleation of bubbles of true vacuum with radii $R\geq R_{\rm c}$ via a dilaton phase transition in the presence of a non-relativistic axion population (light blue). Middle: The expansion of the bubbles compresses the axions and releases energy. The released energy accelerates the axions (blue), which are kinematically forbidden to enter
the dilaton bubbles. Right: The compressed axions exert significant pressure and stabilise the expanding bubble. Axion relic pockets of radius $R_{\rm eq}$ form from regions of false vacuum stabilised by trapped, hot axion gas (dark blue), with an average pocket distance $D$. Figure and text adapted from \cite{Carenza:2024tmi}.}
\label{fig:dark-pockets}        
\end{figure*}

A non-trivial interplay between the two, specifically in the case that the dilaton undergoes a first-order phase transition, can lead to axions trapped in spherical false vacuum regions, separated by a dilaton phase boundary from the surrounding true vacuum. The pressure of the trapped axions can then form stabilised regions of false vacuum, known as Axion Relic Pockets (ARPs) \cite{Carenza:2024tmi} (see also \cite{Witten:1984rs, Hindmarsh:1991ay,Bai:2019, Zhitnitsky_2021, gross2022darkmatterdarkdwarfs} for related models).  An illustration of the formation process is shown in Fig.~\ref{fig:dark-pockets}. At some point after coalescence, the energy density of the pockets redshifts like matter, making them promising candidates for dark matter. Nevertheless, their observational signatures and possible detection strategies are not well understood. The purpose of this paper is to commence the study of ARP phenomenology, focusing in particular on signals sourced by the axion-photon coupling and identifying possible `smoking gun' signals.

ARPs provide a minimal theory for dark matter. For the simplest versions in which the axion is sufficiently weakly coupled to not co-thermalise with the SM, all key predictions of the theory are set by one parameter, which can be taken to be the SM phase transition temperature, $T_t$. Fixing $T_t$ means setting the characteristic pocket radius (to the horizon size at the time of the transition $H^{-1}(T_t)$), the mass and the number \& energy of trapped axions.
The earlier the transition, the smaller \& lighter the pocket and the higher the internal axion energy. 
The transition temperature can cover a wide range of scales.
Gravitational microlensing constraints impose an upper bound on the pocket mass $M/M_\odot<5\cdot 10^{-12}$, which translates to a lower bound on the transition temperature: $T_t\gtrsim 7$ GeV.\footnote{For some benchmark values of $\epsilon^4=0.15\;\alpha=1.6$ as discussed in Sec.~\ref{sec:review}.}
Theoretical consistency enforces an upper bound, $T_t \lesssim 10^{13}$ GeV. 
We will primarily be interested in transition temperatures $T_t\gtrsim 0.5\cdot 10^4\;$TeV, for which the mass and size of the pockets are $M_{\rm pocket}\lesssim 1.4\cdot10^{-29}\;M_\odot = 28.5 {\; \rm kg},\;R_{\rm pocket}\lesssim 1.2\cdot 10^{-11} {\; \rm m}$ and the hot axion gas inside the ARPs, which we assume to be thermalised (though not co-thermalised with the SM), has a  temperature of
$T\gtrsim 2.4\;$TeV. If they comprise dark matter, these pockets are light enough to be locally abundant to the present day, and could give observable signals in terrestrial detectors, through the coupling of the confined axions to SM fields.

In this work, we focus on the well-known coupling of axions to electromagnetism,  
\begin{equation}
    \mathcal{L}_{a\gamma\gamma}=-\frac{1}{4}g_{a\gamma\gamma}\;aF_{\mu\nu}\tilde{F}^{\mu\nu}\;,
\end{equation}
which leads to interconversion of axions and photons in the presence of  background magnetic fields, e.g in astrophysical environments \cite{PhysRevLett.51.1415,Sikivie:1985yu}, or in the presence of a background electric Coulomb field (Primakoff process \cite{PhysRev.81.899,Raffelt:1996wa,Wu_2024}). Inverse Primakoff scattering due to atomic electric fields has been extensively studied and used for experimental searches, e.g.~for solar axions 
\cite{Paschos:1993yf,Avignone_1998,Bernabei:2001ny,Morales_2002,Ahmed_2009,Gao_2020,Dent_2020,Excess_xenon1t,Abe_2021}. Pockets travelling through areas of strong electric or magnetic fields generate a potentially detectable photon flux via inverse Primakoff conversion of the hot axions inside the pockets. However, the calculation of the flux from ARPs differs markedly from the standard calculation. 
 While standard approaches to axion-photon conversion assume asymptotically-plane-wave solutions (both in classical and quantum field theory calculations) \cite{Sikivie:1985yu,PhysRevD.37.1237,Marsh:2021ajy,Sikivie_2021, Wu_2024}, the  axions confined in pockets enjoy reflective boundary conditions on the dilaton wall and do not admit such solutions.
Thus, existing conversion rates and cross-sections for various ALP couplings to photons and matter cannot be straightforwardly used.

We study in detail the conversion of axions trapped inside pockets into photons, in the presence of background electromagnetic fields. Building on the existing literature on axion-photon conversion for asymptotic plane wave axions, we derive new solutions that describe the emitted photon flux while accounting for the finite spherical geometry of the confined axions, both for background electric and magnetic fields. These solutions are the main result of this work, paving the way for phenomenological searches of electromagnetic signals from the pockets both on terrestrial detectors and in astrophysics.

We apply our solutions to search for potential signals of axion relic pockets in different types of terrestrial detectors. We identify upward-going electromagnetic cascades, triggered by axion-photon conversion in atomic electric fields, as a potential smoking-gun signal of the theory. 
For example, we show that the IceCube experiment is sensitive to ARPs passing through the detector volume and interacting with the Coulomb fields of oxygen atoms in the ice. We derive the event rates and place the first experimental constraints on ARPs formed at early cosmic times using the latest data from the IceCube collaboration \cite{Abbasi_2026,abbasi2025improvedmeasurementstevpevextragalactic}. Furthermore, we present rough estimates of the event rate of electromagnetic air showers caused by axion-photon conversion of ARPs in atmospheric detectors (e.g.~the Fluorescence Detector of the Pierre Auger Observatory as well as balloon-borne detectors such as ANITA) and in direct detection experiments (e.g.~XENONnT and LZ). Significantly, IceCube is the strongest probe of ARPs, but only a small region of the parameter space can be consistently probed with current data. For sufficiently strong couplings, the axions co-thermalise with the Standard Model in the early universe, which significantly modifies the scenario of \cite{Carenza:2024tmi}. In this paper, we focus on the original scenario of \cite{Carenza:2024tmi}, and leave cosmological extensions for future work. 

We stress that existing axion constraints do not carry over directly to ARPs. Astrophysical and laboratory searches for axions probe their masses and couplings 
at the \emph{true} vacuum of the dilaton field, but ARP signals are produced by the light, relativistic false-vacuum axions inside the pockets. Indeed, constraints on ARPs are better pictured in the radius-coupling plane, rather than the conventional mass-coupling plane.

This work is structured as follows: in Sec.~\ref{sec:review}, we review the formation and properties of ARPs, following \cite{Carenza:2024tmi}. In Sec.~\ref{sec:conversion} we derive the classical and quantum formulas for axion-photon conversion for axions trapped in spherical pockets. The discussion in this section is quite technical, and readers not interested in all aspects of the calculation can directly skip to the high energy limit of the expression Eq.~\ref{eq:highk}, which has a strikingly similar form with the usual axion photon conversion formula. In Sec.~\ref{sec:atoms}, we reconstruct electric fields inside atoms using numerical fitting formulas for the form factors, and we calculate the average number of emitted photons per passage of an ARP inside an atom. We use our derived expressions to provide the first estimates for the event rates in IceCube in Sec.~\ref{sec:icecube} and for set-ups inspired by ANITA and Pierre Auger in Sec.~\ref{sec:upward}, and we briefly comment on detection prospects for dark matter direct detection experiments. We summarise our results in the conclusion section. The discussion is supplemented with five appendices that include intermediate steps for technical aspects of our work.

\section{Properties of Axion Relic Pockets}
\label{sec:review}
The properties of ARPs are determined by the phase transition that formed them and whether the axion-photon coupling is strong enough to modify the cosmology.
In this section, we first review the main ARP scenario in which the axions that are uncoupled or weakly coupled to photons, following \cite{Carenza:2024tmi}. We then outline how the cosmology is modified when the axions are sufficiently strongly coupled so as to go into thermal equilibrium with the SM plasma in the early universe. We leave the details of this latter scenario to future work.
\subsection{The main scenario: weak axion-photon coupling}
In this scenario, the properties of ARPs are a function of two phase transition parameters; the energy difference between the vacua $\Delta V$, and the wall tension $\sigma$. Fixing the energy density of ARPs to the dark matter density reduces the number of parameters by one, and it's convenient to choose the phase transition temperature, $T_t$, as the remaining free parameter.

For ARPs that account for all of dark matter,  the vacuum energy difference between the true and false vacuum is
\begin{equation}
    \Delta V\Big|_{\rm DM}=19.5\;\text{GeV}^4 \epsilon^{-1}\tilde{g}_{\star }(T_t)\tilde{T}_t^3\;.
\end{equation}
Here, $\epsilon$ parametrises the energy density left in the pockets at a time $t_f$, in which the pockets start to redshift like matter, and is defined as the ratio of scale factors between $t_f$ and at the time of bubble coalescence $t_{\rm coll}$. This number is not a free parameter of the theory and can be constrained by simulations. The parameter can be bounded from above to be $\epsilon^4\lesssim 0.15$. The term $\tilde{g}_{\star}(T_t)=g_{\star}(T_t)/106.75$ represents the number of relativistic degrees of freedom present in the plasma at the time of the transition, while $\tilde{T}_t$ is a rescaled transition temperature $\tilde{T}_t=T_t/1\, \rm{TeV}$.

The equilibrium radius of a stabilised pocket is set by pressure balance and energy conservation to be
\begin{equation}
\label{eq:pocket-radius}
    R_{\rm eq}\equiv R_{\rm pocket}= 2.3\cdot 10^{-4}\, {\rm m} ~(1+\alpha)^{1/3}\, \tilde{g}_\star(T_t)^{-1/2}\, \tilde{T}_t^{-2}\, , 
\end{equation}
where $\alpha$ parametrises the gamma factor of the  walls at coalescence 
\begin{equation}
\gamma_w(t_{\rm coll})=\alpha R_0/R_c\;,
\end{equation}
with $R_0$ the radius of the bubble at $t_{\rm coll}$ and $R_c$ the critical radius. We expect this number to be constrained by the inequality $0<\alpha\lesssim 1.6$. In Sections \ref{sec:icecube}--\ref{sec:upward}, we assume that $\alpha$ saturates the bound $\alpha\simeq1.6$ for simplicity, and leave the determination of this parameter from simulations to future work.

The energy density of a pocket is $4 \Delta V$, where three units of $\Delta V$ comes from the hot axion gas and one unit from the vacuum energy inside the pocket. 
The pocket mass is given by
\begin{equation}
    M_{\rm pocket}=4.9 \cdot 10^{38}\, {\rm GeV}  \left(\frac{1+\alpha}{\epsilon}\right)
     ~\tilde{g}_\star(T_t)^{-1/2}\, \tilde{T}_t^{-3} \, ,
\end{equation}
where the pre-factor in other often-used units is $4.9 \cdot 10^{38}\, {\rm GeV}=8.74\cdot10^{11}\, {\rm kg} =4.39\cdot10^{-19}\, M_\odot$. 

The mean separation of the pockets is given by
\begin{equation}
\label{eq:pocket-distance}
    D_{\rm pocket}=  \left(\frac{M_{\rm pocket}}{\rho_{\rm dm}}\right)^{1/3} =1.1\cdot 10^{11}~{\rm m}\, \left(\frac{1+\alpha}{\epsilon}\right)^{1/3}
    ~\tilde{g}_\star^{-1/6}\, \tilde{T_t}^{-1} \, . 
\end{equation}

The number of hot axions in each pocket is given by
\begin{equation}
\label{eq:axion-number}
    N_{a,\rm pocket}= 3.2\cdot10^{37}\frac{\left(1+\alpha\right)}{\tilde{g}_\star^{3/4}\epsilon^{3/4}}\tilde{T}_t^{-15/4}\;.
\end{equation}

In this paper, we assume that the axion population inside the pocket thermalises through self-interactions (but does not co-thermalise with the hot Big Bang plasma). This makes all predictions of the theory independent of the properties of the seed population of axions prior to the phase transition, and the characteristic axion energy is given by
\begin{equation}
\label{eq:axion-energy}
    E_a\simeq 3 T_a=3\cdot 3.65\;\text{GeV}\left(\frac{\tilde{g}_\star}{\epsilon}\right)^{1/4}\tilde T_t^{3/4}\;,
\end{equation}
where $T_a$ denotes the axion gas temperature. We stress that since the axions inside the pocket are ultra-relativistic, their mass $m_a|_{\rm FV}$ is not a relevant parameter. Constraints on ARP couplings should therefore be expressed in terms of the transition temperature, or an equivalent parameter, e.g.~the ARP mass or radius. 

We close this section by defining the relevant range of scales of the ARPs that  can be detected on Earth. The transition temperature is bounded from below by astrophysical constraints, in particular from microlensing that require $M_{\rm pocket} <5\cdot 10^{-12}\, M_\odot$, which translates into a lower bound on the transition temperature, $\tilde T_t \gtrsim 0.007$. In this low-temperature/high-ARP-mass region, the pockets are approximately meter-sized and weigh as much as a comet. As such, they are much too sparse to frequently interact with terrestrial detectors. 
Thus, in this paper, we focus on a subset of the ARP parameter space in which the pockets are smaller and lighter, for which $\tilde{T}_t\gtrsim 0.5\cdot10^4$. The encounter rate of ARPs with a kilometre-cubed spherical detector volume is of order $\sim 10^{15}\;{\rm yr}^{-1}$ for $\tilde{T}_t=10^{10}$, dropping to $3\times10^{-4}\;{\rm yr}^{-1}$ at $\tilde{T}_t = 0.5 \cdot 10^4$. Still, at the lower temperature end, the encounter rate with Earth remains substantial (approximately $10^5 \;{\rm yr}^{-1}$). % pockets of this kind per year. This motivates considering terrestrial signatures of these objects even beyond the regime in which passages through a single detector are frequent. 
Incidentally, for such transition temperatures the diameters of the pockets satisfy $2R_{\rm pocket} \lesssim 2.4\cdot 10^{-11}\; {\rm m}$ and are small compared to atoms. In Sec.~\ref{sec:atoms}, this will allow us to treat the instantaneous screened atomic electric fields as homogeneous over the pocket volume. The pocket size is also bounded from below. Requiring that $N_{a,\rm pocket}$ is sufficiently large so that the pocket maintains its spherical shape puts an upper limit on the transition temperature, $\tilde T_t \lesssim 10^{10}$. The masses of the pockets in this parameter space range from $28.5\; {\rm kg}$ (for $\tilde T_t = 0.5\cdot 10^4$) to $2\cdot 10^9\; {\rm GeV}$ (for $\tilde T_t =10^{10}$), and contain axions with energies $E_a\simeq 7.3\; {\rm TeV}$ (for $\tilde T_t = 0.5\cdot 10^4$) to $E_a\simeq 0.4\; {\rm EeV}$  (for $\tilde T_t =10^{10}$).

The scenario of \cite{Carenza:2024tmi} was developed for axions and dilatons with no couplings to the Standard Model, and also holds for models with sufficiently small axion-photon coupling. However, if the axion-photon coupling is large enough, the axions co-thermalise with the photons in the hot Big Bang plasma, which drastically alters the cosmology of the model as well as the properties of any axion relic pockets. The dominant interaction between axions and photons is the Primakoff process and its inverse, which have a rate at the ultra-relativistic limit (see e.g. \cite{PhysRevD.33.897, Marsh:2014gca})
\begin{equation}
    \Gamma_{\rm primakoff}(E_a)=n_{\rm ch}(T) \frac{\alpha_{\rm EM} g^2_{a\gamma\gamma}}{8\pi}\ln\left(\frac{4E_a^2}{k^2}\right)\;,
\end{equation}
with $n_{\rm ch}(T)$ the number density of charged particles and $k$ the Debye-H\"{u}ckel inverse radius. Since for the SM plasma, $n_{\rm ch}(T)\sim T^3$ and we expect the logarithm to be an ${\cal O}(1)$ number, we can write $\Gamma_{\rm primakoff}\sim \alpha_{\rm EM} g_{a\gamma\gamma}^2 T^3/8\pi$. Requiring that the axions are not thermalised at the time of the transition requires $\Gamma_{\rm primakoff}/H(T_n)<1$, which translates to an upper bound on the axion to photon coupling. In radiation domination, this consistency relation is
\begin{equation}
\label{eq:consistency-bound}
    |g_{a\gamma\gamma}|\lesssim \left(\frac{\pi^2g_\star(T_n)}{90}\right)^{1/4}\left(\frac{8\pi}{\alpha_{\rm EM} T_n M_{\rm pl}}\right)^{1/2}\;,
\end{equation}
assuming a population of axions created at around the time of the transition, with $M_{\rm pl}$ the reduced Planck mass. In most of this paper, we consider values of $g_{a\gamma\gamma}$ such that Eq.~\eqref{eq:consistency-bound} holds; in section \ref{sec:thermalised}, we briefly discuss how ARP cosmology changes when this equation is not satisfied. 

A non-vanishing axion-photon coupling opens the possibility of axion decay into two photons. Inside the pocket, the false-vacuum axion mass is assumed to be sufficiently small that this process is not relevant. However, across the pocket boundary (of width $\sim m_\phi^{-1}$), the axion mass increases rapidly, which could make `surface emission' of photons possible. Parametrically, stability requires $m_\phi \gtrsim g_{a\gamma\gamma}^2 m_a|_{\rm TV}^3$, which we assume holds in this paper.\footnote{We thank Philip S{\o}rensen for discussions on this point.}

It is worth noting here that a coupling of the axions to the photons suggests a coupling of the dilaton to electromagnetism $g_{\phi\gamma\gamma}$ of comparable magnitude 
\begin{equation}
    \mathcal{L}_{\rm dil}\propto g_{\phi\gamma\gamma}\phi F_{\mu\nu}F^{\mu\nu}\;.
\end{equation} 
This coupling could cause small variations of the fine-structure constant inside the pockets, leading to slightly modified axion-to-photon conversion rates and modified interactions of photons with matter. Constraints on dilatonic couplings \cite{PhysRevLett.134.191003, Baryakhtar:2024rky, Baryakhtar:2025uxs} do not directly apply here, since the variation would only occur in small false vacuum regions and, furthermore,  our dilaton's mass is much heavier than usually considered in the literature. In addition, a coupling of the dilaton to the SM could affect its effective potential in multiple ways. At high temperatures, the scale anomaly \cite{Laine_2010, Saikawa_2018} could induce a non-vanishing vacuum expectation value (vev) for the gauge strength tensor, altering the dynamics of the phase transition. At lower temperatures, we expect background electromagnetic fields to induce a similar vev, but its effect will be suppressed compared to the scales of the dilaton potential. We will assume for the following that these changes are captured within variations of the parameters related to PT described in \cite{Carenza:2024tmi}.

\subsection{Axions strongly coupled to photons}
\label{sec:thermalised}
In this section we briefly discuss the part of the theory space where
\begin{equation}
\label{eq:anti-consistency-bound}
     |g_{a\gamma\gamma}|> \left(\frac{\pi^2g_\star(T_n)}{90}\right)^{1/4}\left(\frac{8\pi}{\alpha_{\rm EM} T_n M_{\rm pl}}\right)^{1/2}\;.
\end{equation}
For such large couplings, the initial axion population is co-thermalised with the SM photon gas at the nucleation temperature $T_n$. In contrast to the original scenario of \cite{Carenza:2024tmi}, this thermalised axion gas exerts sufficient radiation pressure on the nucleated bubbles to block the dilaton phase transition. Energetically, any nucleated true-vacuum bubble must satisfy
$$
\Delta V > {\cal P}(T) + \Delta m_a n_a(T) + \sigma {\cal A}/{\cal V} 
$$
in order to expand. Here, the first term on the right-hand side of the equation encodes the net radiation pressure (${\cal P}(T)\sim T^4$), the second the energy cost of increasing the mass of axions inside the bubble ($\Delta m_a n_a(T) \sim m_a|_{\rm TV} T^3$), and the third the standard bubble wall tension. For $\Delta V \ll T^4$ as in the original scenario of \cite{Carenza:2024tmi}, bubble expansion is blocked for many decades in temperature.    

Eventually, as the universe cools, dilaton bubble expansion becomes possible. We assume that this happens at time $t_e$ (when $T(t_e)=T_e$), which is approximately given by the condition $\Delta V=\mathcal{P}(t_e)$, where we have assumed that the wall tension is negligible and that the massive axions inside the bubble decay away sufficiently quickly to be neglected.  At this point, bubbles nucleate densely (as the temperature-independent nucleation rate satisfies $\Gamma/H(t_e) = H(t_n)/H(t_e) = (T_n/T_e)^2\gg1$). The nucleated bubbles expand slowly and reach terminal velocity almost immediately.

If the axions are sufficiently strongly coupled to remain in thermal equilibrium with the SM plasma at $T_e$ (which corresponds to \eqref{eq:anti-consistency-bound} with $T_n \to T_e$), energy dissipation from the axion gas into the photons is efficient, and the axion radiation pressure cannot increase. Thus, the bubble expansion cannot stop, and pockets cannot form, as long as the axions are in equilibrium with the plasma. 

There are several ways in which the axions can fall out of equilibrium. Geometric decoupling occurs when the false-vacuum regions are of size $R_*\lesssim 1/T_{\rm SM}$, below which the wavelength of the thermal photons exceeds the pocket size. Within the pocket, the total electric and magnetic field is given by the sum over an isotropic distribution of long-wavelength photons, which cancels. However, the number of axions in such small pockets is ${\cal O}(1)$, and classically, one would expect such regions to collapse. Two effects can alter this conclusion: first, consistently characterising such small pockets calls for a full quantum treatment; second, the high curvature of the pockets walls leads to non-thermal axion  production inside the pockets, which may contribute to stabilising the pockets. We postpone a detailed investigation of these possibilities for future work.

For some values of $g_{a\gamma\gamma}$, the axion gas thermally decouples from the SM at some time between $t_n$ and the geometric decoupling time, $t_*$. This leads to complicated phase transition dynamics involving axion shock fronts developing outside the bubbles that may fully or partially thermalise. In that case, stable pockets can form, but their macroscopic properties, like their mass or radius, depend on $g_{a\gamma\gamma}$ through the decoupling time, $t_d$, in addition to the phase transition properties. 
We expect the resulting pockets to be significantly smaller than the Hubble radius at $t_*$. This broadens the range of phenomenologically interesting ARP models, but we leave a detailed exploration of co-thermalised models for future work.

For the rest of this paper, we focus on the original scenario of \cite{Carenza:2024tmi} corresponding to theories satisfying \eqref{eq:consistency-bound}. We note however, that the machinery developed for the phenomenology of the pockets is independent of the scenario adopted, and we expect the methods developed in this to paper extend straightforwardly to models with more strongly coupled axions.

\section{Axion photon conversion inside spherical pockets}
\label{sec:conversion}

In this section, we derive analytical formulas for the axion-photon conversion rate and the emitted power of electromagnetic radiation sourced by the hot axion gas in a background electric field. The spherical symmetry of the pockets and the reflective boundary conditions for the axion field inside the pocket are novel aspects of our setup, and  the resulting axion-photon conversion rate differs from the standard form, which assumes asymptotically free particles admitting plane wave solutions.  

The first key result of this section 
is
Eq.~\eqref{eq:power}, which gives the  emitted power of electromagnetic radiation in a background electric field which does not vary much through the pocket volume, calculated 
in Sec.~\ref{sec:classical} by solving the modified Maxwell equations by calculating an overlap integral over the pocket volume, following a similar treatment as in \cite{Sikivie:1985yu} (for a review see \cite{Sikivie_2021} and references therein). The second key result is Eqs.~\eqref{eq:axionphoton1}-\eqref{eq:highk}, which give the axion-photon conversion rate calculated  
in the quantum theory using Fermi's golden rule. This calculation is  
similar to \cite{PhysRevD.37.1356,Ioannisian_2017}, but adapted to the ARP boundary conditions. The corresponding calculation for the case of constant background magnetic field is given in Appendix~\ref{ap:magn}. These results can be applied to the search of electromagnetic signals from ARPs both for terrestrial and astrophysical searches, assuming the background electromagnetic field is approximately constant in the pocket. We apply our formulas for terrestrial searches of ARPs, where axions are converted into photons due to atomic electric fields.

\subsection{Modified Electrodynamics}
\label{sec:modifiedEM}
We use perturbation theory in $g_{a\gamma\gamma}$ and throughout our calculations use as a building block the solution of the equations of motion for free axions inside a totally reflective sphere
\begin{align}
    &\left(\partial_t^2-\nabla^2+m_a^2\right) a(t,\mathbf{x})=0,\;\\
    & a\left(t,\mathbf{x}=R_{\rm pocket}\hat{\mathbf{n}}\right)=0\;.
\end{align}
The solution is expressed in terms of spherical Bessel functions and spherical harmonics
\begin{align}
\label{eq:free-axion}
    a(t,r,\theta,\phi)=\sum_{n,l,m}\mathcal{C}_{nl}a_{nl}(t) j_{l}(k_{nl}r)Y_{lm}(\theta,\phi)+\text{h.c},
\end{align}
with $\mathcal{C}_{nl}$ the normalisation factor. The momentum $k_{nl}$ is quantized due to the reflective boundary condition
\begin{equation}
    k_{nl}=\frac{\alpha_{nl}}{R_{\rm pocket}}\;,
\end{equation}
with $\alpha_{nl}$ the $n$-th root of the $l$-order spherical Bessel function. The time-dependent part of the equations of motion, provides us with the dispersion relation $\omega_{a,nl}^2=k_{nl}^2+m_a^2\;.$

The interaction between axions and the electromagnetic field
\begin{equation}
    \mathcal{L}_{a\gamma\gamma}=-\frac{1}{4}g_{a\gamma\gamma}a F_{\mu\nu}\tilde{F}^{\mu\nu}=-g_{a\gamma\gamma}a\;\mathbf{E}\cdot\mathbf{B}
\end{equation}
leads to the modified Maxwell's equations \cite{PhysRevLett.58.1799,PhysRevLett.51.1415}
\begin{align}
    &\nabla\cdot\mathbf{E}=\rho-g_{a\gamma\gamma}\mathbf{B}\cdot\nabla a \;,\\
    &\nabla\times\mathbf{B}-\partial_t\mathbf{E}=\mathbf{J}+g_{a\gamma\gamma}\left(\mathbf{B}\partial_t a-\mathbf{E}\times\nabla a\right)
    \;,\\
    &\nabla\times \mathbf{E}+\partial_t\mathbf{B}=0\;,\\
    &\nabla\cdot\mathbf{B}=0\;,\\
    &\left(\partial_t^2-\nabla^2+m_a^2\right)a=g_{a\gamma\gamma}\mathbf{E}\cdot \mathbf{B}\;.
\end{align}
The charge density $\rho$ and current $\mathbf{J}$ refer to external sources that could be present in the system. The interaction between the axion and the electromagnetic fields induces new sources $\rho'= -g_{a\gamma\gamma}\mathbf{B}\cdot\nabla a$ and $\mathbf{J}'=g_{a\gamma\gamma}\left(\mathbf{B}\partial_t a-\mathbf{E}\times\nabla a\right)$ that could lead to the emission of electromagnetic radiation. For the following, we restrict to the lowest order in the coupling $g_{a\gamma\gamma}$, and find radiating solutions given background electric fields that vary slowly inside the pocket.

\subsection{Classical Theory}
\label{sec:classical}
 In this section, we solve the modified Maxwell's equations for an ARP in a homogeneous and time-independent electric field background. The analogous calculation for a background magnetic field is given in Appendix~\ref{ap:magn}. Our method involves linearising the modified Maxwell's equations around the background, and then finding the Poynting vector to determine the power of the emitted radiation, roughly following the procedure in \cite{Sikivie_2021}. In Section \ref{sec:atoms}, we apply these results to subatomic ARPs that travel through atomic electric fields.

The  electromagnetic fields are,  
to first order in perturbations around a background electric field, given by
\begin{align}
   & \mathbf{E}=\mathbf{E}_{0}+\mathbf{e},\\
    &\mathbf{B}=\mathbf{0}+\mathbf{b}\;.
\end{align}
The corresponding equation of motion for the axion then reads
\begin{equation}
    \left(\partial_t^2-\nabla^2+m_a^2\right)a=-g_{a\gamma\gamma}\mathbf{E}_{0}\cdot\mathbf{b}\;,
\end{equation}
in which the source term on the right-hand-side is already second order in perturbation theory. To linear order, the axion field is described by the solutions of the free theory, 
\begin{equation}
    a(r,\theta,\phi,t)=\sum_{nlm}\mathcal{C}_{nl}\;e^{-i\omega_{a,nl}t} j_l (k_{nl}r)Y_{lm}(\theta,\phi)+\text{h.c}\; .
\end{equation}
The equations of the electric and magnetic field are then
\begin{align}
    & \nabla\cdot \mathbf{E}_{0}=\rho\;,\\
    &\left(\partial_t^2-\nabla^2\right)\mathbf{e}=-g_{\rm a\gamma\gamma}\mathbf{E}_{0}\times \nabla \dot{a}\;,\\
    & \left(\partial_t^2-\nabla^2\right)\mathbf{b}=-g_{\rm a\gamma \gamma} \nabla \times \left(\nabla a\times \mathbf{E}_{0}\right)\;.
\end{align}

It is convenient to solve in terms of $A^\mu=(\phi,\mathbf{A)}$ in Lorenz gauge, as the equations then take the familiar form
\begin{align}
&\left(\partial_t^2-\nabla^2\right)\phi=\rho'=0,\\
&\left(\partial_t^2-\nabla^2\right)\mathbf{A}=\mathbf{J}'= g_{a\gamma\gamma}\left(\mathbf{E}_{0}\times\nabla a\right)
\label{eq:EoM}
\end{align}
\paragraph{Single mode} We start discussing the solutions of the above equations for a single axion mode, which we write as
$a(\mathbf{x},t)=e^{-i\omega_{a,nl}t} u_{nlm}(\mathbf{x})$.
Conservation of energy implies a solution for the photon of the form $\mathbf{A}(\mathbf{x},t)=e^{-i\omega_{a,nl}t}\mathbf{A}(\mathbf{x})$, 
and Eq.~\eqref{eq:EoM} simplifies to
\begin{equation}
  \left(-\nabla^2- \omega_{a,nl}^2\right)\mathbf{A}(\mathbf{x})=g_{a\gamma\gamma}\left(\mathbf{E}_{0}\times\nabla u_{nlm}(\mathbf{x})\right)\;.
\end{equation}
This equation can be solved using standard Green's function methods \cite{Jackson:1998nia}
\begin{equation}
    \mathbf{A}(\mathbf{x})=\frac{g_{a\gamma\gamma}}{4\pi}\int_{V_{\rm pocket}}d^3x' \frac{e^{ik_{\gamma
    }|\mathbf{x}-\mathbf{x'}|}}{|\mathbf{x}-\mathbf{x'}|}\left[\mathbf{E}_{0}\times\nabla u_{nlm}(\mathbf{x'})\right]\;,
\end{equation}
with the integral being non-zero only inside the pocket volume $V_{\rm pocket}$. Furthermore, we used the dispersion relation for the photons $k_{\gamma}^2=\omega_{a,nl}^2$. Note that the photon and the axion have same energies $\omega_{a,nl}$, but have different momenta due to the small but non-vanishing axion mass $m_a$.

The relevant regime for emitted radiation is the far field limit, $\mathbf{x}=x\mathbf{\hat{n}}$ with $x\rightarrow\infty$, where the field can be approximated as an outgoing spherical wave
\begin{align}
    \nonumber \mathbf{A}(\mathbf{x})&=\frac{e^{ik_{\gamma}x}g_{a\gamma\gamma}}{4\pi x}\int_{V_{\rm pocket}}d^3x' e^{-ik_{\gamma}\mathbf{\hat{n}}\cdot\mathbf{x}'}\left[\mathbf{E}_{0}\times\nabla u_{nlm}(\mathbf{x'})\right]+\mathcal{O}\left(\frac{1}{x^2}\right) \\ &\equiv\frac{e^{ik_{\gamma}x}g_{a\gamma\gamma}}{4\pi x}\mathbf{F}^{nlm}_{\mathbf{\Omega}}+\mathcal{O}\left(\frac{1}{x^2}\right)\;.
\end{align}
We can then write the time-dependent solution as the sum of the contributions appearing for the free axion in Eq.~\eqref{eq:free-axion}
\begin{equation}
    \mathbf{A}(\mathbf{x},t)=e^{-i\omega_{a,nl}t}\mathbf{A}(\mathbf{x})+e^{i\omega_{a,nl}t}\mathbf{A}^*(\mathbf{x})\;.
\end{equation}
The time-averaged radiated power in direction $\mathbf{\hat{n}}$ sourced by a single $(nlm)$ mode is then defined as
\begin{equation}
    \frac{dP_{nlm}}{d\Omega}=\langle \mathbf{\hat{n}}\cdot \left(\mathbf{E}_{nlm}\times\mathbf{B}_{nlm}\right)\rangle x^2\;.
\end{equation}
The electric and magnetic fields of the radiative solution are given by 
\begin{align}
    \mathbf{B}_{nlm}&=\nabla\times \mathbf{A}(\mathbf x,t)=i\mathbf{k}_\gamma\times \mathbf{A}(\mathbf x,t)+\mathcal{O}(1/x^2)\;,\\
\mathbf{E}_{nlm}&=i\omega_{a,nl}\mathbf{A}(\mathbf{x},t)-i\omega_{a,nl}\left(\hat{\mathbf{k}}_\gamma\cdot\mathbf{A}(\mathbf{x},t)\right)\mathbf{\hat{k}}_\gamma+\mathcal{O}(1/x^2)\;,
\end{align}
where we made use of the Lorenz gauge and defined $\mathbf{k}_\gamma=k_\gamma\mathbf{\hat{n}}$. The second term of the electric field is zero due to the identity $\mathbf{\hat{n}}\cdot\left(\mathbf{\hat{n}}\times\left(\mathbf{\hat{n}}\times\mathbf{A}\right)\right)=0$. Thus, at leading order,
\begin{equation}
\frac{dP_{nlm}}{d\Omega} =\frac{ k_{\gamma}\omega_{a,nl}g_{a\gamma\gamma}^2}{8\pi^2}|\hat{\mathbf{n}}\times \mathbf{F}^{nlm}_{\mathbf{\Omega}}|^2+\mathcal{O}(1/x^2)\;.
\end{equation}
After some manipulations (cf.~Appendix~\ref{app:overlap-integral}), the emitted power for a given $(nlm)$ mode can be expressed as,
\begin{equation}
    \frac{dP_{nlm}}{d\Omega}=\frac{g_{a\gamma\gamma}^2\omega_{a,nl}k_\gamma^3}{8\pi^2 }\mathcal{C}_{nl}^2E^2_{0} \sin^2\psi \Big|\int d^3r e^{-i\mathbf{k_\gamma} \mathbf{r}} u_{nlm}(\mathbf r)\Big|^2\;,
\end{equation}
with $\hat{\mathbf{k}}_\gamma\cdot \mathbf{E}_{0}=E_{0}\cos\psi$, and we assumed the background field is approximately constant within the pocket volume.
The above overlap integral can be analytically solved (cf.~Appendix \ref{app:overlap-integral}), and  results in
\begin{align}
    \frac{dP_{nlm}}{d\Omega}=2k_\gamma\omega_{a,nl}g_{a\gamma\gamma}^2E^2_{0}\mathcal{C}_{nl}^2\sin^2\psi |Y_{lm}(\mathbf{\hat{k}_\gamma})|^2\frac{F_{nl}^2(R_{\rm pocket})}{m_a^4}\;,
\end{align}
with
\begin{equation}
   F_{nl}\left(R_{\rm pocket}\right)=R_{\rm pocket}^2k_\gamma k_{nl} \;j_{l-1}\left(k_{nl}R_{\rm pocket}\right)j_l(k_\gamma R_{\rm pocket})\;,
\end{equation}
where we have used that $k_\gamma^2-k_{nl}^2=m_a^2$. Note that the expression is finite for $m_a\rightarrow0$, since in that limit $j_l(k_\gamma R_{\rm pocket})\rightarrow0$ in the numerator.

The angular integral can be solved by converting $\sin^2\psi$ to  spherical harmonics and then using the Wigner-3j symbol \cite{Varshalovich:1988ifq}. Parametrising the direction of the photon momentum as $$\mathbf{\hat{n}}=(\sin\psi\sin\phi,\sin\psi\cos\phi,\cos\psi) \, ,$$ 
we have
\begin{equation}    P_{nlm}=\frac{4k_\gamma\omega_{a,nl}g_{a\gamma\gamma}^2E^2_{0}\mathcal{C}_{nl}^2F_{nl}^2(R_{\rm pocket})}{3m_a^4}\left(1-\frac{l(l+1)-3m^2}{(2l-1)(2l+3)}\right)\;,
\end{equation}
which is our final expression for the power of the emitted radiation from a single axion mode. 
\paragraph{Sum over modes}Since we are dealing with a classical calculation, it is useful to extend our result to a sum over possible $(nlm)$ modes. Then, the radiating solution takes the form
\begin{equation}
    \mathbf{A}(\mathbf{x})=\frac{e^{ik_{\gamma}x}}{4\pi x}\sum_{nlm}\mathbf{F}^{nlm}_{\mathbf\Omega}\;.
\end{equation}
Extending the Poynting vector calculation, we have that
\begin{equation}
    \frac{dP}{d\Omega}=\sum_{nlm}\sum_{n'l'm'}\langle \mathbf{\hat{n}}\cdot\left(\mathbf{E}_{nlm}\times \mathbf{B}_{n'l'm'}\right)\rangle x^2\;,
\end{equation}
where the emitted power is still calculated in direction $\mathbf{k}_\gamma=k_\gamma\hat{\mathbf{n}}$. The time average of the Poynting vector enforces $\omega_{a,nl}= \omega_{n'l'}$, which eliminates all cross-terms between unequal primed and unprimed indices for $n$ and $l$. 
Then we are left with a degeneracy on $m$ states, so averaging over them we have
\begin{equation}
    \frac{dP}{d\Omega}=\frac{1}{2l+1}\sum_{n,l,m,m'}\frac{\omega_{a,nl}k_\gamma g_{a\gamma\gamma}^2}{8\pi^2}\left(\mathbf{F}^{nlm}_{\mathbf{\Omega}}\cdot \mathbf{F}^{nlm'}_{\mathbf{\Omega}}-\left(\hat{\mathbf{n}}\cdot\mathbf{F}^{nlm}_{\mathbf{\Omega}}\right)\left(\hat{\mathbf{n}}\cdot\mathbf{F}^{nlm'}_{\mathbf{\Omega}}\right)\right)\;.
\end{equation}
The kernel $\mathbf{F}^{nlm}_{\mathbf{\Omega}}$ has a directional dependence proportional to $\mathbf{k}_\gamma\times\mathbf{E}_{0}$ (see App.~\ref{app:overlap-integral} for more details on this), so the total emitted power per solid angle can be further simplified 
\begin{equation}
    \frac{dP}{d\Omega}=\frac{1}{2l+1}\sum_{n,l,m,m'}2k_\gamma\omega_{a,nl}g_{a\gamma\gamma}^2\mathcal{C}_{nl}^2E^2_{\rm atom}\sin^2\psi Y_{lm}(\mathbf{\hat{k}_\gamma})Y^*_{lm'}(\mathbf{\hat{k}_\gamma})\frac{F_{nl}^2(R_{\rm pocket})}{m_a^4}\;.
\end{equation}
The integral over the solid angle can be done analytically as in the single mode case; however, here the Wigner-3j symbol enforces the selection rule $m=m'$. The total emitted power is then
\begin{equation}
    \label{eq:power}
    P=\sum_{n,l}\frac{4k_\gamma\omega_{a,nl}g_{a\gamma\gamma}^2\mathcal{C}_{nl}^2E^2_{0}F_{nl}^2(R_{\rm pocket})}{3m_a^4}\;.
\end{equation}
This equation is our final result for the instantaneous emission from an ARP in a constant electric field. In Sec.~\ref{sec:atoms}, we apply these results to an ARP traversing an atom, and in Sec.~\ref{sec:icecube}, to a dark matter population of ARPs travelling through a medium.

\subsection{Quantum Theory}
\label{sec:quantum}

So far, our treatment of the axion-photon conversion has been fully classical. However, a realistic description of the process involves the conversion of single quanta of the axion and photon fields in the presence of a background electric field. This motivates a quantum description, which we here develop.

In this section, we calculate the axion-photon-mixing 
matrix element by expanding the fields into the corresponding mode functions and using time-dependent perturbation theory. We derive the axion-photon conversion rate using Fermi's golden rule, and find that the tree-level result matches the classical result of Sec.~\ref{sec:classical}. 

We leave the investigation of the general ARP quantum field theory, including also interactions to other fields, to future work.

We expand the axion field into modes of the spherical pocket
\begin{align}
   &a(\mathbf{r},t)=\sum_{n,l,m}\frac{\mathcal{B}_{nl}}{\sqrt{2\omega_{a,nl}}}\left(\hat{a}_{nlm} u_{nlm}(\mathbf{r})e^{-i\omega_{a,nl}t}+\hat{a}^\dagger_{nlm} u^*_{nlm}(\mathbf{r})e^{i\omega_{a,nl}t}\right)\;,
\end{align}
with $u_{nlm}=j_l(k_{nl}r)Y_{lm}(\mathbf{\hat{r}})$, and $\mathcal{B}_{nl}$ the corresponding normalisation, which in contrast to the classical case, does not involve the energy factor $\sqrt{1/2\omega_{a,nl}}$. The two normalisations are related by $\mathcal{C}_{nl}=\sqrt{N/2\omega_{a,nl}}\mathcal{B}_{nl}$, with $N$ the occupation number for each state in the classical case. The axion field is expanded into creation and annihilation operators of an $(nlm)$ state such that $\hat{a}_{nlm}|0\rangle=0$ and $\hat{a}^\dagger_{nlm}|0\rangle=|nlm\rangle$.

The photons do not feel the pocket boundaries at all, so their corresponding mode functions are still plane waves. For convenience, we use a discrete summation over the mode functions inside a fiducial volume $V$

\begin{equation}
    \mathbf{A}=\sum_\mathbf{k}\frac{1}{\sqrt{2\omega_\mathbf{\gamma,k}V}}\left(\sum_\lambda\left( \hat{A}_\mathbf{k}^\lambda \vec{\mathbf{\epsilon}}^\lambda(k)e^{-i(\omega_\gamma t-\mathbf{k\mathbf{x})}}+\hat{A}_\mathbf{k}^{\dagger\lambda}\vec{\mathbf{\epsilon}}^{*\lambda}(k)e^{+i(\omega_\gamma t-\mathbf{k\mathbf{x})}}\right)\right)\;,
\end{equation}
The matrix element for axion-photon conversion is then given by
\begin{equation}
\mathcal{M}=-i\langle \gamma|\mathcal{H}_{\rm int}|a\rangle\;,
\end{equation}
with
\begin{equation}
    \mathcal{H}_{\rm int}=-g_{a\gamma\gamma}a\mathbf{E}\cdot \mathbf{B}.
\end{equation}
We expand the corresponding Hamiltonian in the presence of a background electric field $\mathcal{H}_{\rm int}=-g_{a\gamma\gamma}a\mathbf{E}_{0}\cdot \mathbf{B}$, thus the dynamical quantity corresponding to the emitted photon field is then $\mathbf{B}=\nabla\times \mathbf{A}$. 

The matrix element for a conversion of a given $(nlm)$ mode has the form $\langle\mathbf{k}_\gamma|\mathcal{H}_{\rm int}|nlm\rangle$. We observe that the only contributing terms from the mode expansions involve the $\hat{A}^\dagger_{\mathbf k_\gamma}$ mode for the photon and the $\hat{a}_{nlm}$ for the axion, so
\begin{align}
&\mathcal{M}=\frac{g_{a\gamma\gamma}\mathcal{B}_{nl}}{2\omega_{\gamma} \sqrt{V}}(\mathbf{k}_\gamma\times \vec{\epsilon}^{*,\lambda})\cdot \mathbf{E}_{0}\int d^3r e^{-i\mathbf{k_\gamma} \mathbf{r}} u_{nlm}(\mathbf r)\;.
\end{align}
We note that the corresponding integral is only over the space argument $d^3r$, omitting the energy conserving time integral $2\pi \delta(\omega_{a,nl}-\omega_{\gamma})$ which will explicitly appear in the expression of Fermi's golden rule.

For the final expression, we need to sum over the photon polarisations, so using  $\sum_\lambda {\epsilon_i}\epsilon^*_j=\delta_{ij}$, the final squared amplitude can be expressed as an overlap integral in the pocket volume
\begin{equation}
    \sum_\lambda |\mathcal M|^2=\left(\frac{g_{a\gamma\gamma}}{2\omega_{\gamma} }\right)^2\frac{\mathcal{B}_{nl}^2}{V} E_{0}^2 k_\gamma^2 \sin^2\psi \Big|\int d^3r e^{-i\mathbf{k_\gamma} \mathbf{r}} u_{nlm}(\mathbf r)\Big|^2\;.
\end{equation}
The overlap integral we need to calculate, is identical to the one we encountered in the classical case. So now, given this squared amplitude, we calculate the axion-photon conversion rate via Fermi's golden rule
\begin{equation}
    \Gamma_{a\rightarrow \gamma}=2\pi  \sum_\mathbf{k}\sum_\lambda |\mathcal M|^2 \delta(\omega_{a,nl}-\omega_{\gamma})\;,
\end{equation}
and by converting to the continuum again, using $\sum_{\mathbf{k}}\rightarrow V\omega^2\int d\Omega_{\mathbf{k}}/(2\pi)^3$ with $\omega=\omega_{a,nl}=\omega_{\gamma}$, we obtain
\begin{equation}
    \Gamma_{a\rightarrow \gamma}=\frac{g_{a\gamma\gamma}^2\omega^2}{16\pi^2 }\mathcal{B}_{nl}^2E^2_{0}\int d\Omega_k \sin^2\psi \Big|\int d^3r e^{-i\mathbf{k_\gamma} \mathbf{r}} u_{nlm}(\mathbf r)\Big|^2\;.
\end{equation}
Performing the integral as in the classical case, we arrive at
\begin{equation}
    \Gamma_{a\rightarrow \gamma}=\frac{2}{3m_a^4}g_{a\gamma\gamma}^2E^2_{0}\mathcal{B}^2_{nl}F_{nl}^2\left(R_{\rm pocket}\right)\left(1-\frac{l(l+1)-3m^2}{(2l-1)(2l+3)}\right)\;,
\end{equation}

We observe that, as expected, the result for the interaction of a single mode exactly reproduces the classical case, since the emitted power is related to the conversion rate via $P_{nlm}=N\omega_{a,nl}\Gamma_{a\rightarrow\gamma}$.

Specifying the axion energy, $E_a$, fixes the momentum $k_a$ and the quantum numbers $n$ and $l$, but there is still a degeneracy of different $m$-states. 

Here, we consider 
axions with definite $(nl)$ numbers only, and average over $m$ to obtain the  squared amplitude 
\begin{equation}
|\mathcal{M}|^2_{\rm avg}=\frac{1}{2l+1}\sum_m|\mathcal{M}|^2\;.
\end{equation} 
The corresponding axion-photon conversion rate has the simple form
\begin{equation}
\label{eq:axionphoton1}
    \Gamma_{a\rightarrow \gamma}^{nl}=\frac{2}{3m_a^4}g_{a\gamma\gamma}^2E^2_{0}\mathcal{B}^2_{nl}F_{nl}^2\left(R_{\rm pocket}\right)\;.
\end{equation}
A regime of particular importance for our scenarios is for $k_aR\gg1$, in which the axion-photon conversion rate significantly simplifies: 
\begin{equation}
\label{eq:highk}
 \Gamma_{a\rightarrow \gamma}^{nl}=\frac{g_{a\gamma\gamma}^2 E^2_{0} R_{\rm pocket}}{3}\;.
\end{equation}
We note that this result is independent of energy, as in the usual plane-wave calculation in the small-mixing limit,  as reviewed in App~\ref{ap:plane-waves}.  We refer to App.~\ref{app:overlap-integral} for more details on this derivation.

We would like to highlight again, that these equations describe the instantaneous conversion rate 
for a constant electric field. We will now turn the time and space-dependent situation of an ARP traversing an atomic electric field.

\section{Axion relic pockets inside atoms}
\label{sec:atoms}

ARPs that are sufficiently light to be locally abundant today could enter detectors at Earth and generate detectable signals. For trapped axions coupled to electromagnetism, the primary interaction mechanism is conversion into photons in the presence of a background electric field. Since the radius of such a pocket is much smaller than the size of atoms, we expect the dominant contribution to the axion–photon conversion rate to arise from atomic electric fields.

It is well known that atomic electric fields differ from the simple Coulomb law, due to screening effects that depend on the electron number density of the atom. This screening effect is encoded in Fourier space through an atomic form factor. Such form factors have been extensively studied in atomic physics and can be computed with high precision using the Hartree–Fock approximation\footnote{We note that for atoms with higher atomic number e.g.~xenon, the approximation does not include correlation effects between electrons, and as a result fails to reproduce all features of the atomic structure \cite{Liu:2024bhy}. We expect, however, that the approximation gives a good estimate for the energy ranges we consider.} \cite{Coulthard_1967}. Tabulated results for these calculations exist for all elements, along with analytic fitting formulas valid over specific ranges of momentum transfer between the axion and photon \cite{Brown2004}. The range of validity of these fits also determines the spatial resolution we can achieve when using them for the reconstruction of the atomic electric fields.

Recent studies have incorporated realistic atomic form factors in the context of inverse Primakoff scattering of solar and supernova axions \cite{Gao_2020, Abe_2021}. We follow the approach of \cite{Abe_2021}, which uses a fitted analytic function of the form factor, derived from  relativistic Hartree–Fock approximations as presented in \cite{Brown2004}. However, for our calculation we need to reconstruct the full atomic electric field $\mathbf{E}_{\rm atom}$ as function of the distance from the atomic centre. 
\subsection{Reconstruction of electric fields}
The static electric field of the atom satisfies the Poisson equation
\begin{equation}
    \nabla\cdot\mathbf{E}_{\rm atom}=\rho(\mathbf{x})\;.
\end{equation}
It is evident that to determine the form of the electric field, we need to have an understanding of the charge density inside the atom.
The charge density consists of the Z number of protons in the atomic centre, and electrons with a number density $n_e(\mathbf{x})$, thus
\begin{equation}
    \rho(\mathbf{x})=Z_{\rm atom}e\delta^3(\mathbf{x})-en_e(\mathbf{x)}.\\
\end{equation}
The form factor is defined as the Fourier transform of the electron number density \cite{Abe_2021}
\begin{equation}
    F(\mathbf{q})=\int d^3x e^{-i\mathbf{q}\mathbf{x}}n_e(\mathbf{x)}\;,
\end{equation}
so, provided that we have a formula for $F(\mathbf{q})$, we can reverse the Fourier transform to obtain an expression for $n_e(\mathbf{x})$
\begin{equation}
   n_e(\mathbf{x}) =\frac{1}{(2\pi)^3}\int d^3q e^{i\mathbf{qx}}F(q)\;.
\end{equation}
There exist different parametrisations of $F(\mathbf{q})$, as shown in  Appendix~\ref{ap:form-factors}, but here we 
focus on
\begin{equation}
    F(q)=\sum_{i=1}^4a_i \exp\left[-b_i\left(\frac{|\mathbf{q}|}{4\pi}\right)^2\right]+c,\;
\end{equation}
where the constants $(a_i,b_i,c)$ are different for each atom. We provide the specific numbers for oxygen and nitrogen in Appendix~\ref{ap:form-factors}. This expression is accurate up to $q\lesssim q_{\rm max}= 25 \;\mathring A^{-1}$. 

The electron number density will then be
\begin{equation}
    n_e(r)=\frac{1}{2\pi^2}\int dq\; \frac{q\sin(qr) F (q)}{r}= c\frac{\delta(r)}{4\pi r^2}+8\pi^{3/2}\sum_i \frac{a_i}{b_i^{3/2}}e^{-4\pi r^2/b_i}\;,
\end{equation}
where the analytical form is meant to be trusted for $r\gtrsim r_{min}\sim 1/q_{\rm max}$. For $q\gtrsim 25 \mathring{A}$ one can use different fitting formulas (see e.g. \cite{Brown2004}) and eventually approach the regime where the atom is unscreened $F(q)\to 0$.

Knowing the electron number density, we are now in a position to solve the continuity equation of the electric field. Observing that the charge density $\rho$ is spherically symmetric in this approximation, we have
\begin{equation}
    \mathbf{E}_{\rm atom}(r)=\frac{e}{4\pi}\frac{Z-N_e(r)}{r^2}\hat{\mathbf{r}},\; 
\end{equation}
with
\begin{equation}
    N_e(r)=\int_0^r dr' 4\pi r'^2 n_e(r')\;.
\end{equation}

As mentioned before, for $r\lesssim r_{\rm min}$ the approximation ceases to be valid. To incorporate effects for $r\ll r_{\rm min}$ we use 
\begin{align}
\mathbf{E}_{\rm atom}(r)=
\begin{cases}
    \frac{e}{4\pi}\frac{Z_{\rm atom}-N_e(r)}{r^2}\hat{\mathbf{r}}\;,\; r>r_* \\ \frac{e}{4\pi}\frac{Z_{\rm atom}}{r^2}\hat{\mathbf{r}} \;,\; r\lesssim r_*\;,
\end{cases}
\end{align}
with $r_*$ the radius at which the two expressions coincide, which is at the same order of magnitude as $r_{\rm min}$. A more precise treatment would require introducing an interpolating region between the two regimes, where a different fit would be more appropriate. However, we do not expect this to significantly affect the remainder of our calculations.

\begin{figure*}[t]
    \centering
    \includegraphics[width=0.65\textwidth]{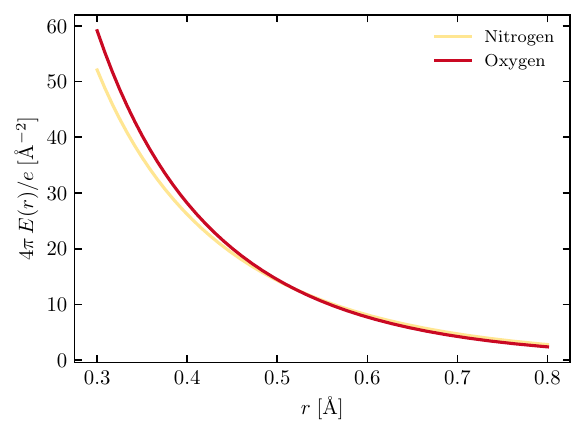}
    \caption{ The reconstructed electric field for oxygen and nitrogen.}
\label{fig:Electric-fields}        
\end{figure*}

The reconstructed electric fields for oxygen and nitrogen are shown in Fig.~\ref{fig:Electric-fields} and the corresponding values for the fit can be found in App.~\ref{ap:form-factors}.

\subsection{Emitted photons per ARP passage}
\begin{figure*}[t]
    \centering
    \includegraphics[width=0.9\textwidth]{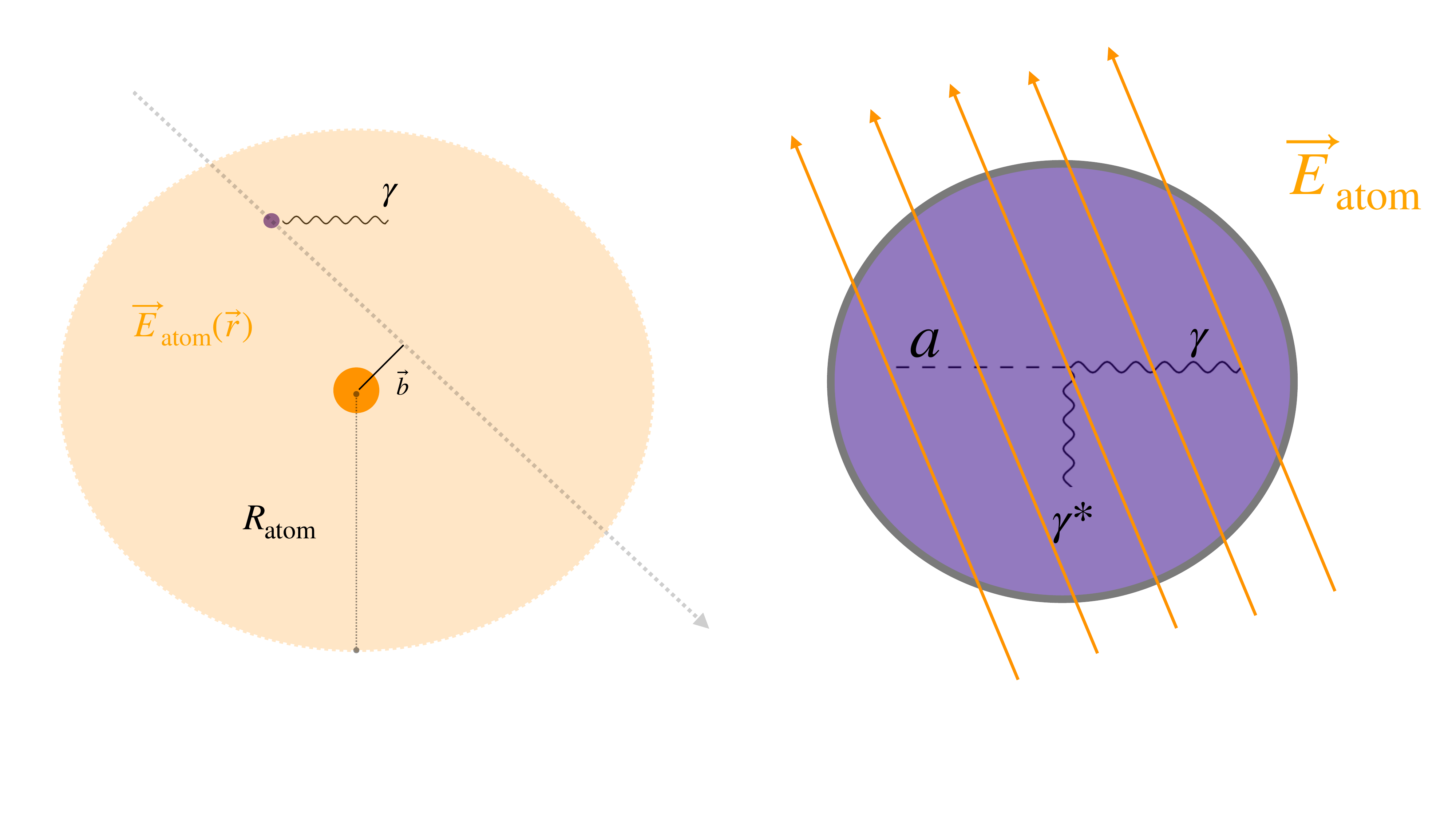}
    \caption{\textbf{Left}: Illustration of an ARP passing through an atom, travelling along a chord with impact parameter b from the atomic centre. The orange blob in the centre depicts the nucleus, while the fainter orange background represents the electron cloud. The whole structure induces an atomic electric $\mathbf{E}_{\rm atom}$, reconstructed via the atomic form factors, and assuming spherical symmetry. The purple dot is the axion relic pocket. \textbf{Right}: A zoomed-in illustration towards the ARP. As it passes through the atoms, the pocket experiences a background atomic electric field, which is approximately constant inside its radius. The trapped axions inside convert into the photons in the presence of the background electric field. For non-relativistic ARPs, the distribution of the emitted photons is isotropic.}
\label{fig:ARP-passage}        
\end{figure*}

As the ARP enters and travels inside the atom, it will experience a varying electric field, depending on its distance from the atomic centre. For ARPs that are small with respect to the atom, and move at non-relativistic speeds, the electric field is homogeneous inside the pocket and its variation is slow compared to the crossing time of the hot axions inside the pocket. This motivates the approximation of an instantaneously constant and homogeneous electric field. As the ARP traverses the atom, a number of photons are emitted from axion-photon conversion. An illustration of the process is shown schematically in Fig.~\ref{fig:ARP-passage}. The instantaneous rate  of emitted photons depends on the ARP trajectory through the atom and is given by
\begin{equation}
\label{eq:photonrate}
    \dot{N}_\gamma= \Gamma_{a\rightarrow \gamma} N_a,
\end{equation}
with $N_a$ the number of axions inside the pocket, and $\Gamma_{a\rightarrow\gamma}$ the axion-photon conversion rate, which for a definite $(n,l)$ state is
\begin{equation}
    \Gamma_{a\rightarrow \gamma}=\frac{2}{3m_a^4}g_{a\gamma\gamma}^2E^2_{\rm atom}\left(t\right)\mathcal{B}^2_{nl}F_{nl}^2\left(R_{\rm pocket}\right)\;.
\end{equation}
The time dependence of the atomic field is inherited by the motion of the ARP in the atom, since $E_{\rm atom}=E_{\rm atom}(d)$, with $d=d(t)$ the distance of the ARP from the atomic centre. Since, the rest of the conversion rate does not depend on time, we can use a shorthand notation of the form $\Gamma_{a\rightarrow\gamma}=\mathcal{A}(R_{\rm pocket})E^2_{\rm atom}(t)$.

Given this form for the conversion rate, we can find the total emitted photons per passage of an ARP inside an atom (assuming the number of axions remains approximately the same) by integrating over time Eq.~\eqref{eq:photonrate}
\begin{equation}
    N_{\gamma,\rm pass}=N_a \mathcal{A}(R_{\rm pocket}) \int_{t_{\rm enter}}^{t_{\rm exit}}dt E_{\rm atom}^2(t)\;,
    \label{eq:Ngamma1}
\end{equation}
with $t_{\rm enter},\;t_{\rm exit}$ the times when the ARP enters and exits the atom respectively. The ARP travels in a straight line within the atom, so we can parametrise its trajectory using chords with impact parameter, $b$, and a straight-line parametrisation
\begin{equation}
    d(t)=\sqrt{b^2+v^2_{\rm pocket}t^2}\;,
\end{equation}
such that $d(t_{\rm min})=b$ for $t_{\rm min}=0$ and $d(t_{\rm exit})= R_{\rm atom}$ for $t_{\rm exit}^2=(R_{\rm atom}^2-b^2)/v_{\rm pocket}^2$. We take the ARP velocity to be the dark matter virial velocity in the Milky Way, $v_{\rm pocket}=v_{\rm DM}\simeq 10^{-3}c$. Evaluating the integral \eqref{eq:Ngamma1} then gives the number of emitted photons for an ARP crossing an atom with a given impact parameter, $N_{\gamma,\rm pass}=N_{\gamma,\rm pass}(b)$. Considering the average over many atom crossings, the impact parameter is uniformly distributed in the transverse area, and the mean number of emitted photons per atom crossing is,
\begin{equation}
 \langle  N_{\gamma,\rm pass}\rangle = \frac{ \int_{b_{\rm min}}^{R_{\rm atom}}db\; 2\pi b\;    N_{\gamma,\rm pass}(b)}{ \int_{b_{\rm min}}^{R_{\rm atom}}db\; 2\pi b  }=\int_{b_{\rm min}}^{R_{\rm atom}}db \frac{2b}{R^2_{\rm atom}-b_{\rm min}^2} N_{\gamma,\rm pass}(b)\;.
\end{equation}
The lower cut-off $b_{\rm min}$ of the impact parameter is set by the size of the pocket and by the regime of validity of the approximations used in deriving the conversion rate. For $R_{\rm pocket}\lesssim R_{\rm nucleus}$, the atomic-field description is physically cut off at approximately the nuclear radius, $b_{\rm min}\simeq R_{\rm nucleus}$. For larger pockets, $R_{\rm pocket}\gtrsim R_{\rm nucleus}$, the relevant cut-off is instead controlled by the pocket size, and we parametrise it as $ b_{\rm min}\simeq \xi R_{\rm pocket}$ where $\xi=\mathcal O(1)$. The value of $\xi$ controls the accuracy of the homogeneous-field treatment in the pocket. Spatial variations of the electric field across the pocket induce gradient corrections that, due to spherical symmetry, scale parametrically with pocket distance $d$, as $(R_{\rm pocket}/d)^2$ after performing the angular integrations. In the remainder of this paper, we adopt $\xi=3/2$ as a representative choice, which keeps the centre of the pocket separated from the nucleus by more than one pocket radius while accounting for the close encounters that tend to dominate the conversion rate. In total
\begin{align}
\label{eq:bmin}
    b_{\rm min}=\begin{cases} R_{{\rm nucleus}}\;,\; \xi R_{\rm pocket}(T)\lesssim R_{\rm nucleus}\;\\
    \xi R_{\rm pocket}(T)\;,\; \xi R_{\rm pocket}(T)\gtrsim R_{\rm nucleus}\;.
    \end{cases}
\end{align}

After performing the double integration (cf.~App.~\ref{ap:event-rate}), we find for the average number of emitted photons
\begin{equation}
    \langle  N_{\gamma,\rm pass}\rangle =\frac{8 N_a}{3v_{\rm DM}m_a^4}g_{a\gamma\gamma}^2 \mathcal{B}_{nl}^2\frac{F_{nl}^2(R_{\rm pocket})}{R_{\rm atom}^2-b_{\rm min}^2}\int dr r\sqrt{r^2-b_{\rm min}^2}E^2_{\rm atom}(r)\;.
    \label{eq:Ngammaav}
\end{equation}
For the high energy limit $k_a R_{\rm pocket}\gg1$, the equation is then reduced to
\begin{equation}
\label{eq:Ngammaav-highk}
    \langle N_{\gamma,\rm pass}\rangle=\frac{4N_a}{3\left(R_{\rm atom}^2-b_{\rm min}^2\right)}\frac{g_{a\gamma\gamma}^2R_{\rm pocket}}{v_{\rm DM}}\int_{b_{\rm min}}^{R_{\rm atom} }dr\; r \sqrt{r^2-b_{\rm min}^2} E^2_{\rm atom}(r).
\end{equation}
These two equations provide the necessary ingredients for calculating the possible signals of the pocket.

Knowing the average number of photons emitted per ARP passage inside an atom, we are in a position to estimate event rates that could be detected in terrestrial detectors which are sensitive to detecting electromagnetic radiation. We will provide these estimates and set the first constraints on the ARP scenario for different examples of terrestrial detectors in the next section.

\section{Detecting ARPs with IceCube}
\label{sec:icecube}

The signal from ARPs is not delicate or subtle; it is an electromagnetic  `loud bang' as the emitted high-energy photon cascades into a shower of photons, electrons, and positrons. Expressed in terms of the phase transition temperature, the relevant ARP parameter range that can be constrained by terrestrial detectors is $0.5\cdot 10^4\; {\rm TeV} \lesssim T_t \lesssim 10^{10}\; {\rm TeV}$, where the lower bound consistently enforces that the pockets are sufficiently subatomic for our approximations to work. 
The relevant characteristic energies of the trapped axions, which also sets the energy of photons emitted through axion-photon conversion ($\omega_\gamma=E_a$), then range between $7.3\; {\rm TeV} \lesssim E_a \lesssim 0.4\; {\rm EeV}$, cf.~\eqref{eq:axion-energy}. 

The emitted photons promptly interact with their environment. The dominant interaction of such high-energy photons is the Bethe-Heitler process, $\gamma + N \to e^++ e^-+N$, in which the photon undergoes electron-positron pair production from scattering off nuclear electric field. Considering ice, for example, the mean free path of the primary photon is around $0.5$ meters at TeV energies, growing to $30$ meters at EeV energies due to the Landau–Pomeranchuk–Migdal effect.  Subsequent interactions of the secondaries give rise to an electromagnetic cascade and associated Cherenkov radiation. The longitudinal length of the EM shower in the ice is expected to range from a few metres at TeV energies to ${\cal O}(100)$ meters at the highest energies considered.

IceCube is sensitive to the ARP signal: its large detector volume provides a large effective cross-section for ARP interactions, and its 11-year dataset provides significant exposure. The large volume of the detector ensures that typical ARP events originate and are contained in the bulk of the detector, where their signal is an excess of electron scattering events \cite{Cui_2022}.\footnote{We note that the energies considered here are far above the $100\;$GeV detector threshold of IceCube
, and the $\mathcal{O}(10)$GeV threshold of the DeepCore sub-array.
} The primary backgrounds, including electron-neutrino initiated charge-current cascades, differ significantly in their angular distribution from the isotropic ARP signal.

A `smoking gun' signal of ARP dark matter is upward-going electromagnetic showers uniformly distributed over the detector volume. The primary photons generated from axion-photon conversion of the ultrarelativistic axion gas are isotropically distributed in the ARP rest frame, and since the ARP moves non-relativistically, the photon distribution is highly isotropic also in the lab frame. Half the events will therefore involve upward-going, high-energy photons. This is interesting, as the upward-going flux of atmospheric neutrinos becomes increasingly attenuated by the Earth's absorption for energies $E_{\rm event } \gtrsim 40$ TeV \cite{IceCube:2017roe}. For $100\; {\rm TeV} \lesssim E_{\rm event} \lesssim 10^3\; {\rm TeV}$, the dominant background is expected to be diffuse astrophysical flux, and possibly yet-to-be-observed prompt atmospheric flux produced from the decay of charmed hadrons \cite{Abbasi_2021,yuan2023updateddirectionsicecubehese,IceCube:2024fxo}. In the PeV--EeV range, also the astrophysical neutrino background rapidly decreases, potentially providing a clean detection channel for ARPs. The spectrum of the ARP signal is set by the distribution of the hot axions inside the pocket, and would stand out as a `bump' above the falling power-law backgrounds.

In the current data, there is no sign of ARPs. No anomaly consistent with the ARP predictions has yet been reported by IceCube, and in this paper, we compute the first constraints on ARPs by applying and extending the results from Sec.~\ref{sec:atoms}
to the reported IceCube spectrum of events from the full $11.4$-year data set
at energies $E_{\rm event}\gtrsim 5\;\rm{TeV}$  \cite{Abbasi_2026,abbasi2025improvedmeasurementstevpevextragalactic}.

A total of $4,949$ ``Medium Energy Starting Events" (MESE) cascades were recorded in the energy range $5\;\rm{TeV}\lesssim E\lesssim 10\;\rm{PeV}$, with the highest reconstructed energy of $E_{\rm max}\simeq1.5\;\rm{PeV}$. 
For energies above $E\gtrsim10\;\rm{PeV}$,  no events are reported. 
The number of events per energy bin is given in Fig.~1 of \cite{Abbasi_2026}, and the data for each bin, along with the expected SM background from a Monte Carlo simulation, are found in the dataset \cite{DVN/ZBO52I_2026}.

\subsection{ARP event rates} 

To determine how IceCube constrains the ARP theory, we assume, for simplicity, that the detector is completely efficient at detecting ARP-generated events. A dedicated analysis would account for the detector efficiency; however, the EM cascades from ARPs have high energies and are typically contained within the detector volume, $V_{\rm det}$, and we expect the efficiency to be high. Moreover, we neglect ARP-generated showers originating outside of, but extending into, the detector volume.

The expected number of ARP-generated events can be calculated as follows. Assuming dark matter consists of ARPs, the flux of pockets into the detector volume is expected to be approximately isotropic. Each pocket will travel along a straight chord through the detector, and axion-photon conversion is possible in the electric fields of the water molecules in the ice, which we approximate as oxygen atoms. We determine the number of emitted photons per unit time by multiplying the number of ARP-encounters per unit time (i.e.~ARP flux times the detector area) with the average number of atoms encountered on a path through the detector and the average number of emitted photons per atom (cf.~Eq.~\eqref{eq:Ngammaav}).

For an average ARP trajectory of chord length $\langle L\rangle$ in the detector, the total number of photons produced is
\begin{equation}
     N_{\rm chord}=n_{\rm atom} \sigma_{\rm atom}\;\langle L\rangle \; \langle  N_{\gamma,\rm pass}\rangle\;,
\end{equation}
with $\sigma_{\rm atom}$ the atomic cross-section for the atom-ARP interaction and $n_{\rm atom}$ the oxygen number density.
The isotropic ARP flux is given by $\Phi=n_{\rm pocket}v_{\rm DM}/4$, which gives a total event rate of
\begin{equation}
\label{eq:event-rate1}
 \dot{N}_{\rm events}=\Phi_{\rm pocket} A_{\rm det} N_{\rm chord}\;.
\end{equation}

For thermalised ARPs, $T_a R_{\rm pocket} \gg 1$, and the high-energy limit ($k_aR_{\rm pocket}\gg1$) of Eq.~\eqref{eq:highk} is always applicable. As shown in App.~\ref{ap:event-rate}, this results in a rather simple expression for the event rate:
\begin{equation}
\dot{N}_{\rm events}=g_{a\gamma\gamma}^2 \frac{\alpha_{\rm EM}  N_O R_{\rm pocket}N_a}{3D^3 }\int_{b_{\rm min}}^{R_{\rm atom}} dr \;r\sqrt{r^2-b_{\rm min}^2} \left(\frac{Z_{\rm atom}-N_e(r)}{r^2}\right)^2\;,
\label{eq:event-rate2}
\end{equation}
with $\alpha_{\rm EM}$ the fine structure constant, $Z_O=8$ the atomic number of oxygen, $N_O$ the number of oxygen atoms in the detector and $R_{\rm atom}$ the atomic radius of oxygen. The number of axions in the pocket $N_a$, its radius $R_{\rm pocket}$ and average pocket distance $D$ are given by Eqs.~\eqref{eq:axion-number},    \eqref{eq:pocket-radius}, \eqref{eq:pocket-distance}, respectively. 

Since the interaction rate in the high-energy limit is energy-independent (cf.~Eq.~\eqref{eq:highk}), the spectrum of the emitted photons is identical to the axion spectrum; for a thermalised gas, it follows the Planck distribution with temperature $T_a$. For simplicity, we here further simplify the analysis by approximating the gas as mono-energetic with energy set by Eq.~\eqref{eq:axion-energy}. 

The event rate \eqref{eq:event-rate2} has a power-law dependence on 
the transition temperature, $\dot{N}_{\rm event}\sim T_t^{-\beta}$, with $\beta$ a positive constant, for $R_{\rm pocket}\lesssim 2/3R_{\rm nucleus}=2\;{\rm fm}$. For $R_{\rm pocket}\gtrsim 2/3R_{\rm nucleus}$, the lower cut-off of the impact parameter $b_{\rm min}$ depends on $R_{\rm pocket}$ changing the slope of the event rate. Finally for larger pockets, the integral involving atomic electric fields is mostly over the region where the fields are screened. As a result, the integral is more suppressed, which causes the event rate to decrease. Models 
in which the phase transition occurred at early times leads to numerous pockets filled with highly energetic axions; however, in these models, the event rate drops because they contain fewer axions and have smaller radii.  

Here, we present a simplified analysis illustrating how the ARP theory can be constrained by IceCube. We use the latest IceCube MESE events reported in \cite{Abbasi_2026,DVN/ZBO52I_2026} to derive a limit on the axion-photon coupling, under a few simplifying assumptions. The cascade data of \cite{Abbasi_2026,DVN/ZBO52I_2026} is well-fitted 
by the combined `background' of atmospheric and astrophysical neutrinos, in addition to muons. We use the residuals of this fit to derive a limit on a mono-energetic ARP signal. A more detailed analysis would account for the spectral shape of the ARP signal and simultaneously fit the background and signal. However, as the IceCube energy bins are sparse in the range we are interested in, and there is no excess in the data, we expect that our analysis provides a good approximation of the full analysis. To derive Poisson upper limits on the signal, we use the Helene prescription, which conditions on a positive signal \cite{Helene:1982pb}. 
The 90\% confidence limit on the signal, $N_s$, is calculated by determining the values of $\lambda$ that satisfy 

\begin{equation}
   \frac{1}{\cal D} \sum_{k=0}^{N_{\rm obs}}\frac{\lambda^ke^{-\lambda}}{k!}\leq0.1\;,
\end{equation}
where ${\cal D} =\sum_{l=0}^{N_{\rm obs}}\frac{N_{\rm MC}^l e^{-N_{\rm MC}}}{l!}$
with $\lambda=N_{\rm MC}+N^{\rm up}_{\rm s}$. Here, $N_{\rm obs},\;N_{\rm MC}$ correspond to the observed events and estimated background from \cite{Abbasi_2026},
 while $N^{\rm up}_{\rm s}$ denotes the upper bound on the signal. For the energy range where IceCube has not observed any events (and the MC background is low), this prescription yields $N_s^{\rm up} \lesssim 2.3$ at 90\% c.l..

\begin{figure*}[t]
    \centering
    \includegraphics[width=0.9\textwidth]{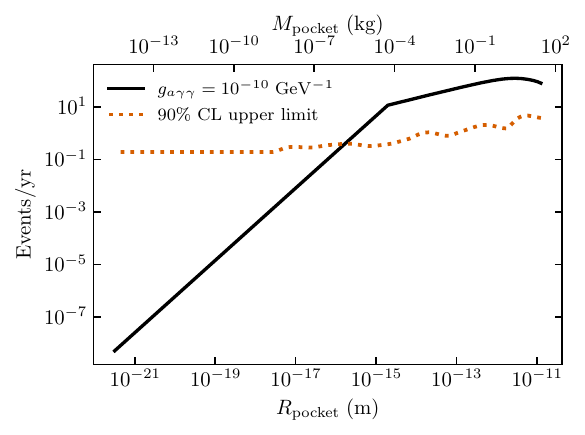}
    \caption{Expected event rate per year as a function of pocket radius in IceCube for $g_{\rm a\gamma\gamma}=10^{-10} \;\rm GeV^{-1}$ and a choice of parameters $\alpha=1.6$,
$\epsilon^4=0.15$,
$N_O=3.1\times10^{37},\; \; R_{\rm atom}=1.52\mathring{A}.$ The dotted red line corresponds to the event rate obtained by dividing the upper limit, $N_s^{\rm up}$, by the exposure time of the detector, as described in the main text.}
\label{fig:ICECUBE1}        
\end{figure*}

\begin{figure*}[t]
    \centering
    \includegraphics[width=0.9\textwidth]{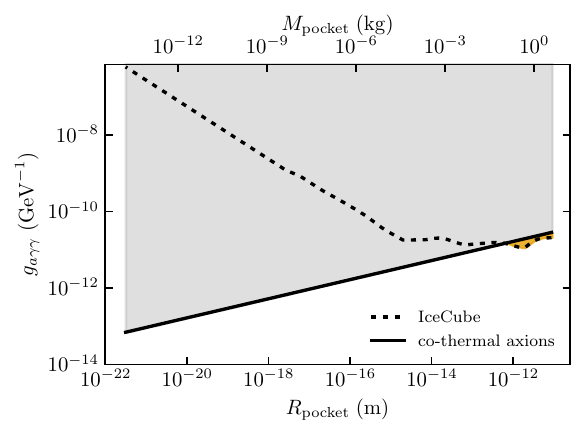}
    \caption{ Constraint on the axion-photon coupling as a function of pocket radius, assuming a $90\%$ confidence limit and the Helene prescription. We focus on models in which the axions are sufficiently weakly coupled not to co-thermalise with the SM, which fall below the black solid line. The parameter space ruled out by IceCube MESE data is marked in orange. The extrapolation of the IceCube constraint into the gray region above the co-thermalisation constraint is for illustration purposes only. 
    }
\label{fig:ICECUBE2}        
\end{figure*}

Figure~\ref{fig:ICECUBE1} shows the expected event rate in IceCube as a function of the pocket radius for a fixed axion-photon coupling of $g_{a\gamma\gamma}\simeq 10^{-10}\;\rm GeV^{-1}$ (arbitrarily chosen for illustration purposes). The dotted line corresponds to the upper bound on the signal $N_s^{\rm up}$ divided by the exposure time of the detector. As mentioned above, smaller pockets contain fewer high-energy axions and generate fewer signal events, even if the flux of pockets through the detector is larger. The slope of the event rate changes with the size of the pockets, due to the different $b_{\rm min}$ (c.f. Eq.~\ref{eq:bmin}), as discussed previously. 
Figure~\ref{fig:ICECUBE2} shows our 90\% confidence-interval upper limit on the axion-photon coupling for ARP dark matter. Equation \eqref{eq:consistency-bound} cuts straight through the parameter space: in this paper, we focus on the simple scenario of \cite{Carenza:2024tmi} rather than the more involved versions discussed in section \ref{sec:thermalised}, and hence, we can only consistently constrain models with parameters \emph{below} the black solid line. Among these, IceCube constrains models with quite massive pockets ($M \sim {\cal O}({\rm kg})$) of almost atomic size -- these are the first experimental constraints on ARP dark matter. The IceCube constraint applies to models with phase transition temperatures of $0.5\cdot 10^{4}\; {\rm TeV} \lesssim T_t \lesssim 8\cdot10^{7}\; {\rm TeV}$.\footnote{Thus, there is a small high-$T_t$ region ($8\cdot 10^{7}\; {\rm TeV} \lesssim T_t \lesssim 10^{9}\; {\rm TeV}$) and a larger small-$T_t$ region ($7\cdot 10^{-3}\; {\rm TeV} \lesssim T_t \lesssim 0.5 \cdot 10^{4}\; {\rm TeV}$) that remain untouched by this IceCube bound.} 
For illustration purposes, we extrapolate the IceCube constraint to smaller and lighter pockets. Larger detector volumes and longer exposure times could probe more of the ARP parameter space.


\section{Other terrestrial probes of axion relic pockets}
\label{sec:upward}
The ARP signal is quite simple: an isotropic event rate of high-energy electromagnetic cascades that scales with the number of atoms in the detector. Neutrino and cosmic ray detectors as well as low-background dark matter direct detection experiments can, in principle, be sensitive to this signal. Here, we schematically discuss the prospects for detecting ARPs with two types of detectors: atmospheric air-shower detectors and underground, dense-target direct-detection experiments. 

\subsection{Atmospheric detectors}
Any object approaching Earth must pass through its atmosphere, and sufficiently light dark matter candidates flow through it all the time. ARPs interacting with atomic Coulomb fields in the atmosphere generate `flashes' of electromagnetic radiation. A primary photon of TeV energy has a mean-free path of $\sim 390$ meters at sea level (growing with altitude as the atmospheric density decreases), and triggers an electromagnetic cascade with a longitudinal extent of $\sim6$ kilometers. At EeV energies, the mean-free path is $\sim 820$ meters at sea level, and the shower length grows to $\sim 10$ kilometers. 

Air-shower detectors, such as the Pierre Auger Observatory \cite{Abraham_2010} and the balloon-borne ANITA \cite{Gorham_2009,Gorham_2010,Gorham_2018,Gorham_2019} and PUEO \cite{Abarr_2021} payloads, leverage the Earth's atmosphere as a vast
observatory and are sensitive to 
extensive particle cascades, which are observed via a complementary set of signatures, including secondary particles at ground level, nitrogen fluorescence, and coherent radio emission. 
These observatories are sensitive to the downward-going cosmic-ray flux and the atmospheric and astrophysical neutrino flux, which grows increasingly anisotropic at energies above a few TeV as the Earth's neutrino opacity decreases.  
At sufficiently high energies, there is no Standard Model process that generates upward-going electromagnetic cascades in directions well below the horizon.
Thus, similar to the case of IceCube, upward-going atmospheric electromagnetic cascades with  TeV to EeV energies provide a smoking-gun prediction of ARP dark matter.

The goal of this section is to calculate the rate of ARP-induced  events that atmospheric detectors could  try to detect. We consider two cases of atmospheric detectors, a ground based fluorescence detector modelled on the Pierre Auger Observatory and a balloon-borne detector modelled on ANITA. The derived event rates related to the ground-based detector closely resemble IceCube, with the main difference being a height dependent number density of the atoms in the atmosphere. By contrast, determining the detector-volume geometry for balloon-borne detectors requires accounting for both the detector's field of view and the Cherenkov cone of the primary particle.

\subsubsection{Balloon-borne detectors}
Floating high in the atmosphere, balloon-borne detectors are sensitive to events across an enormous volume.
In this section, we calculate an idealised event rate of upward-going atmospheric electromagnetic cascades from ARPs for a hypothetical balloon-borne payload with parameters similar to ANITA. For any given experiment, our results in this section can be simply modified and paired with instrument-specific factors, such as the exposure and detector acceptance, to determine the expected observed event rate.

The balloon-borne detectors we consider fly at an altitude $h$ and look down towards Earth. The `detection rate' of  showers from ARPs is given by
\begin{equation}
\dot{N}_{\rm obs} = \int dV \int d\Omega \frac{d^2 n_\gamma(E_a, {\bf r},  \dot {\bf  r})}{d t d\Omega} \cdot P_{\rm obs} \, , 
\end{equation}
where the volume integral runs over the observed atmospheric sub-horizon volume, and the solid angle integral runs over the direction of the emitted photon, ensuring that the detector is within the light cone of the extended air shower (EAS). The number of photons of energy $E$ emitted from ARPs per unit time and solid angle is denoted by  $\frac{d^2 n_\gamma(E_a, {\bf r}, {\bf \dot r})}{d t d\Omega}$. Finally, $P_{\rm obs}$ denotes the experiment-dependent acceptance, encoding the probability that an air shower directed towards the detector from a primary photon with energy $E$ emitted at position ${\bf r}$ is recorded as an event. Determining $P_{\rm obs}$ involves simulating the peak electric field at the location of the detector, and the probability of an event being triggered, which depends on the beam pattern of antennas. Correctly characterising $P_{\rm obs}$ is necessary to translate theoretically predicted photon fluxes into an expected number of detected events. Here, we instead consider an idealised event rate including all EAS directed towards the detector, similar to e.g.~\cite{Motloch:2013kva}. This rate takes a simple analytic form, and provides insights into the relative merits of different detection strategies. The idealised event rate is given by
\begin{equation}
    \dot{N} = \int dV \int d\Omega  \frac{d^2 n_\gamma(E_a, {\bf r}, \dot {\bf r})}{d t d\Omega}  \, .
    \label{eq:ideal-rate}
\end{equation}
The distribution of primary photons is isotropic and given by 
\begin{equation}
    \frac{d^2 n_\gamma(E_a, {\bf r}, \dot {\bf r})}{d t d\Omega}  = \frac{1}{4\pi}n_{\rm pocket} N_a \Gamma_{\rm eff} \, ,
    \label{eq:dn2dEdOmega}
\end{equation}
where $n_{\rm pocket}$ denotes the number density of axion relic pockets, $N_a$ the number of axions in each pocket, and $\Gamma_{\rm eff}$ the effective rate of the $a\to \gamma$ process, which includes contributions from all species of particles present in the atmosphere (oxygen and nitrogen, etc.), $\Gamma_{\rm eff} = \sum_i \Gamma^i_{\rm eff}$. The rate of photon-emission for a single ARP passing through the atmosphere consisting of target atom species $i$ with number density $n_i$ is given by the average number of photons emitted per passage of an ARP within a single such atom, $\langle N^i_{\gamma,\rm pass}\rangle$, times the encounter rate between an ARP and the corresponding atoms, $v_{\rm pocket}\sigma_i n_{i}(z)$, where $\sigma_i$ is the atomic cross-section. Combined, this gives, 
\begin{equation}
    N_a\Gamma^i_{\rm eff}= \langle N^i_{\gamma,\rm pass}\rangle v_{\rm pocket}\sigma_i n_{i}(z)\;.
\end{equation}

Accounting for all species,  the photon emission rate per ARP passing through the atmosphere is $N_a \Gamma_{\rm eff}  = \sum_i \langle N^i_{\gamma,\rm pass}\rangle v_{\rm pocket}\sigma_i n_{i}(z)$, where  $z=r-R_E$ for the Earth radius $R_E =6,357\; {\rm km}$. As the number density of target particles decreases with altitude, so does the emission rate. We model the target particle number density as $n_i(r)= \niz {\rm exp}(- (r-R_E)/H_s)$, where $H_s=7.6\; {\rm km}$ denotes the scale height, which we take to be species-independent. 

For an ARP signal to be detectable, the photon must be emitted in the direction of the balloon.
Air showers within the observable, sub-horizon volume can, in principle, be detected if the detector is within the opening angle of the Cherenkov cone, $\theta_d(z)$. The Cherenkov cone narrows with altitude, as the atmosphere becomes increasingly thin. 
The Cherenkov cone opening angle depends on the frequency of the signal and the density of the atmosphere as $\cos \theta_d = 1/(\beta n)$, where $\beta$ denotes the speed of the primary particle, $\beta=v/c$, and $n$ denotes the index of refraction, which is, in general, frequency-dependent. Since the EAS are initiated by ultra-relativistic particles, $\beta=1$. Moreover, in the atmosphere, $n=1 + \delta n$ with $\delta  n \ll 1$, so that $\cos \theta_d = 1/(1 + \delta n)= 1 - \delta n$. Consequently,  $\Delta \Omega(r) = 2\pi \delta n$. Since the deviation of the index of refraction from unity, $\delta n$, is proportional to the atmospheric density of particles, which we take to be exponential:
$
n_{\rm atm}(r) = n_{\rm atm}^0 e^{- (r-R_E)/H_s} \, ,
$
the radially-dependent solid angle is given by
\begin{equation}
\Delta \Omega(r) =\DOz ~{\rm exp}\left[- (r-R_E)/H_s\right] \, .
\end{equation}

\begin{figure*}[t]
    \centering
    \includegraphics[width=0.8\textwidth]{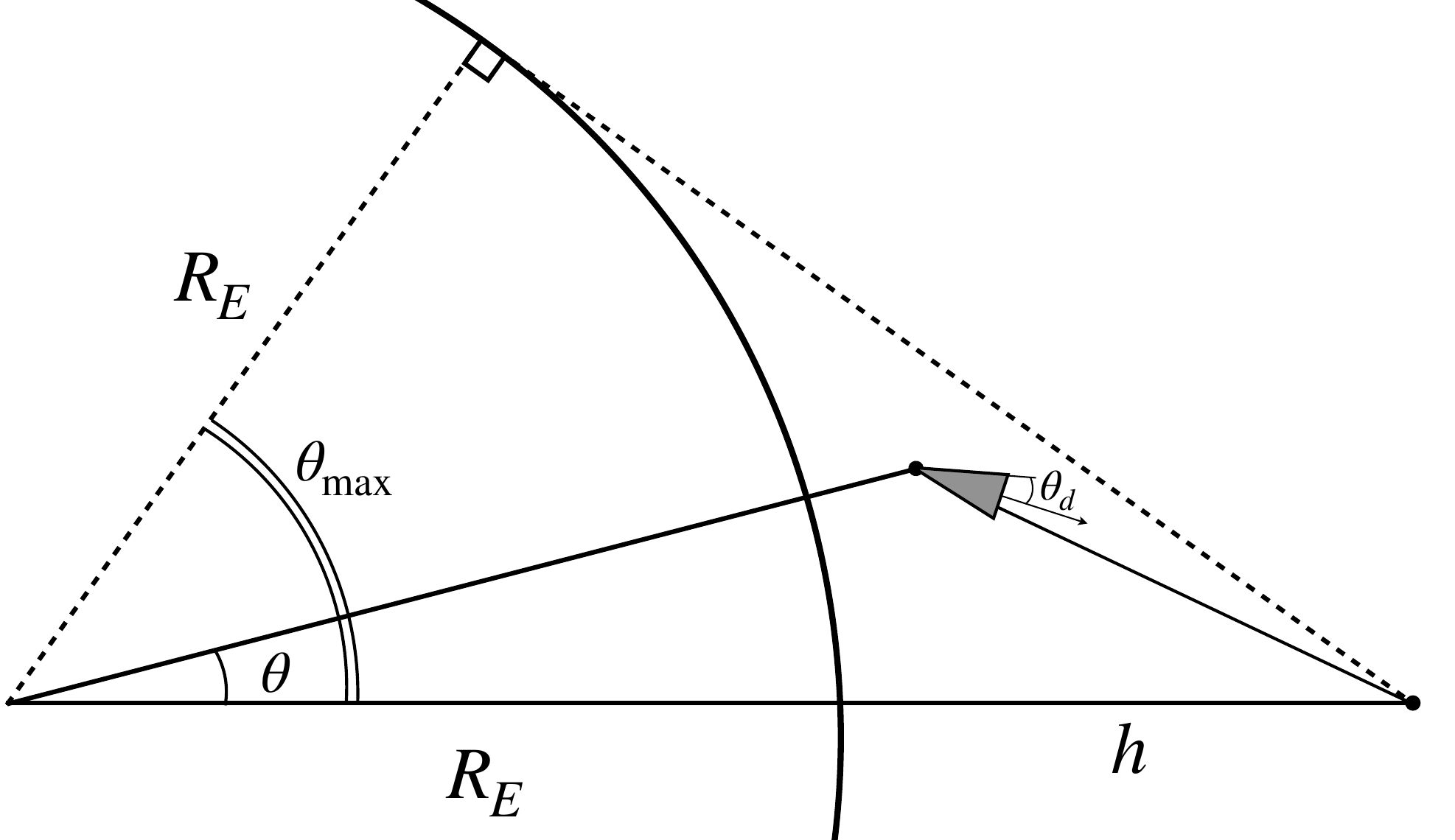}
    \caption{A sketch illustrating the geometry of the ARP induced cascades in the atmosphere and the observability of a balloon-borne detector. $R_E$ corresponds to the Earth's radius, and h to the height of the detector. ARPs interact with atomic electric fields and convert to photons, which subsequently induce electromagnetic cascades. The Cherenkov opening angle is defined as $\theta_d$. The angle $\theta$ corresponds to the angular position of the emission point compared to the detector, while $\theta_{\rm max}$ corresponds to the maximum angle that is visible from the balloon.}
    \label{fig:ANITAgeom}
\end{figure*}

The idealised event rate is therefore given by
\begin{equation}
   \dot{N} = 
  \frac{1}{4}  n_{\rm pocket}   v_{\rm pocket} \sum_i \left( \langle N^i_{\gamma,\rm pass}\rangle \sigma_{i} n_{i}^{(0)}  \right) \DOz
   \int dV \;  e^{-2\left(\frac{r-R_E}{H_s}\right)} \, .
\end{equation}
The geometry of the volume integral is shown in Fig.~\ref{fig:ANITAgeom}. Roughly $78.1\%$ of the atmosphere consists of nitrogen molecules; thus, as a first approximation, we consider only nitrogen in our calculation. One can account for all available species by including the rest of the constituents, but we do not expect this to affect our conclusions. After some algebra presented in detail in App.~\ref{ap:anita}, the event rate as a function of the height of the balloon is
\begin{align} 
\notag&\dot{N}(h) =\frac{\alpha\pi g_{\rm a\gamma\gamma}^2}{12}\frac{N_aR_{\rm pocket}}{D^3}N_{N,0}(h)\Delta\Omega^{0}\\&\times \int_{b_{\rm min}}^{R_{\rm atom}} dr \;r\sqrt{r^2-b_{\rm min}^2} \left(\frac{Z_{\rm atom}-N_e(r)}{r^2}\right)^2\;,
\label{eq:bbd-rate}
\end{align}
with $\Delta\Omega^0=2\pi\delta n^0=1.9\times 10^{-3}$ and $N_{N,0}(h)=n_{N,0}R_EH_s\left(2h+H_s\right)$, and for the height of the balloon $h\simeq 37$ km and $H=7.6$ km, the species scale. The subscript ``N'' refers to nitrogen-related quantities, i.e.~$N_{N}(h),\;n_{N,0}$ respectively denote the number of nitrogen atoms at a balloon height h,  and their number density at the surface of the Earth, while $Z_N=7$ is the nitrogen atomic number. We stress that the competitive edge of balloon-borne payloads lies in the enormous atmospheric volume that the detector can be sensitive to: $N_{N, 0}(h) \sim R_E H_s h$. Eq.~\ref{eq:bbd-rate} is the main result of this section, and is similar to Eq.~\ref{eq:event-rate2} for the IceCube experiment. %, with the main difference being the height-dependent number of nitrogen atoms that are included in the detector's field of view. 

The expected ARP-signal event rate  is shown in Fig.~\ref{fig:all} as a function of transition temperature.\footnote{The ANITA payload flew at a height of 35 -- 37 kilometres above the Antarctic ice and was able to observe a surface area of approximately $1.5
\;\rm M\;km^2$ \cite{Hoover_2010}.} For pockets smaller than the nitrogen nucleus size, the event rate exhibits again a power-law behaviour with respect to the transition temperature $R\sim T_t^{-\beta}$, with $\beta$ a positive number. The slope changes as in the IceCube case for larger pockets and eventually, the integral of the electric field suppresses the event rate. This implies that smaller (and hence more abundant) ARPs, which formed at earlier times, will have suppressed event rates compared to larger pockets. Furthermore, the event rate increases with the height of the balloon, as more area is covered by the detector. We stress that observatories like ANITA are only sensitive to the smallest ARPs with the highest internal axion energies (towards the left end of Fig.~\ref{fig:all}).
For these, the event rates are highly suppressed in the ARP models we consider, and no observable signal is expected. Indeed, the ARP theories considered in this paper cannot explain the `anomalous' upward-going ANITA events with reconstructed shower energies of $0.6\pm0.4$ EeV and $0.56^{+0.3}_{-0.2}$ EeV, respectively \cite{PhysRevLett.117.071101,PhysRevLett.126.071103}.\footnote{
The Pierre Auger Observatory has
used their fluorescence detector capabilities to search for upward-going air showers \cite{PierreAuger:2023elf}, finding no candidate events and deriving strong constraints on the rate.
}

\subsubsection{Pierre Auger Observatory}
The Fluorescence Detector (FD) of the Pierre Auger Observatory (PAO) consists of 24 telescopes that detect air showers via the fluorescent light emitted isotropically by atmospheric nitrogen. The light is collected by the telescopes into a camera consisting of photomultiplier tubes (PMT), with pixel field of view $\sim1.5^\circ$. Each telescope has a field of view $30^\circ\times30^\circ$ in azimuth and elevation \cite{PierreAuger:2015eyc}.  Due to its long observation time and wide field of view, it is also sensitive to upward-going air showers. As the cascade progresses it illuminates successive pixels at different times. The time sequence of pixel activation is used to determine the directionality of the shower \cite{PierreAuger:2023elf}. As mentioned above, PAO has not observed upward-going events compatible with the  ANITA anomalies \cite{PierreAuger:2021gci, PierreAuger:2023elf}, which places strong constraints on most models trying to explain those events. Below, we discuss how to interpret these results in the context of ARPs.

In this section, we derive a first, crude estimate of the ARP event rates that could ideally be detected by the FD detector of PAO. Similarly to the case of ANITA, we assume that the detector could detect all events happening inside its effective volume. The effective volume of the FD is complicated as it depends on the geographic locations of the telescopes, their viewing angle, and the energies of the events. Here, we consider a simplified model of the relevant volume by   
assuming it covers a surface area of $S=10^4\;\rm{km}^2$, and 
up to a maximum height for the first interaction $H=9\;\rm{km}$. These numbers correspond to the ones used for simulated events in \cite{PierreAuger:2021gci, PhysRevLett.134.121003}.

As above, we model the density of atoms in the atmosphere as an exponential $n_i=n_{i,0}\exp{(-z/H_s)}$ with $H_s=7.6$km. The idealised event rate is again given by Eq.~\ref{eq:ideal-rate}.

Focussing on the dominant, nitrogen component, we find 

\begin{align}
  &\notag \dot{N} =\frac{2\alpha g_{\rm a\gamma\gamma}^2}{3} \frac{N_aR_{\rm pocket}}{D^3} N_{N,0}(1-e^{-H/H_s}) \\&\times\int_{b_{\rm min}}^{R_{\rm atom}} dr \;r\sqrt{r^2-b_{\rm min}^2} \left(\frac{Z_{\rm atom}-N_e(r)}{r^2}\right)^2\;,
\end{align}
where we defined $N_{N,0}=n_{N,0} V_{\rm atm}S_{\rm det}H_s$. See App.~\ref{ap:PA} for more details.
We show the annual event rate in Fig.~\ref{fig:all}, with the signal exhibiting the usual behaviour with respect to the transition temperature. The event rate of pockets that formed in earlier times is again suppressed due to the smaller number of confined axions and smaller size.
We find that the idealised event rates for PAO are higher than those of balloon-borne payloads, but we stress that the rates are not strictly identical and this conclusion could change once further detector-specific considerations are accounted for. In particular, our estimates of the event rate within the field-of-view of balloon-borne payloads account for the directionality of the EAS. By contrast, our estimate for PAO counts \emph{all} ARP-induced events within the field-of-view of the FD, including both up-going and down-going showers. Deriving constraints on ARPs from PAO requires accounting for relevant backgrounds and the exposure. However, we expect PAO to be at least as sensitive to ARP-induced signals as balloon-borne payloads. In particular, we note that most upward-going EAS generated by ARPs initiate at low altitudes, where the atmospheric density is the highest. Thus, the PAO analysis of the anomalous ANITA events in \cite{PierreAuger:2021gci, PierreAuger:2023elf} carries over to ARPs: we expect that if the anomalous ANITA events were explained by ARPs, the large-exposure PAO should have seen similar events as well.\footnote{As discussed in section \ref{sec:thermalised}, our analysis is limited to sufficiently weakly coupled ARPs that never co-thermalise with the SM, as the cosmology is more involved for more strongly coupled ARPs. However, we expect these conclusions to carry over to more strongly coupled ARPs, as the morphology of the ARP-induced atmospheric signal primarily depends on the atmospheric density.} However, as was the case for ANITA, the detector's threshold to very energetic showers limits the parameter space towards the smaller pockets. As a result, weakly coupled ARPs might not be observable, due to their suppressed event rates.

%%%%%%%%%%%%%%%%%%%%%%%%%%%%%%%%%%%%%%%%%%%%%%%%%%%%%
\subsection{Direct detection experiments}
In this section, we briefly discuss the prospects of searching for ARPs through axion-photon conversion in  low-background direct detection experiments. The high energy of the emitted primary photons is a characteristic signal, distinct from the more traditional signals from electron and nuclear recoil of other dark matter candidates that these experiments were designed to search for. 

\subsubsection{Liquid Noble-Gas Detectors}
%%%%%%%%%%%%%%%%%%%%%%%%%%%%%%%%%%%%%%%%%%%%%%%%%%%%%
Direct detection experiments based on liquefied noble gases are of great interest for detecting dark matter particles. These experiments have increased greatly in sensitivity over the years, e.g~by roughly four orders of magnitude over the past $\sim 18$ years \cite{XENON10:2008low, LZ:2024zvo}. Noble gases are good scintillators with low ionisation energies, and both liquid Argon (LAr) and liquid xenon (LXe) have been used as detector material, both in the liquid-only phase and as dual-phase detectors. Examples of argon-based detectors include DarkSide50 \cite{Wright_2012}, DEAP-3600 \cite{Lai:2023qub}, and the forthcoming DarkSide-20k \cite{Zani:2024ybb}. Xenon-based experiments include PandaX-4T \cite{PandaX-4T:2021bab}, XENONnT \cite{XENON:2024wpa}, LUX-ZEPLIN \cite{LZ:2019sgr}, XMASS \cite{Abe:2013tc},  and the planned upgrades PandaX-xT \cite{PANDA-X:2024dlo} and XLZD \cite{XLZD_2025}. Several of these experiments have published searches for conventional, massive dark matter axions; however, these searches are often deep within the regions excluded by astrophysical bounds \cite{Ferreira:2022egk}. 

The ARP signal from axion-photon conversion in the detector volume consists of a high-energy photon primary in the TeV -- EeV range. Focussing on the denser, liquid phase of these detectors (in which the conversion rate is larger), such a primary has a mean free path of $\sim 4\; $cm at TeV energies, growing to $\sim8\; $m at EeV energies.

Direct detection experiments can be sensitive to relativistic axions produced in the sun, which, e.g., may be detected by the axio-electric effect if the axion couples to electrons. In this paper, we do not assume direct couplings to electrons but consider axion-photon conversion in the atomic Coulomb fields of the detector material, as in the previous sections.

\begin{figure*}[t]
    \centering
    \includegraphics[width=0.8\textwidth]{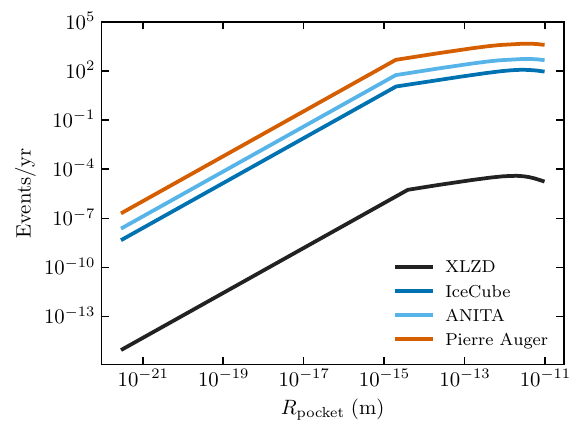}
    \caption{Event rate per year comparison for PAO, ANITA, XLZD and IceCube, with $g_{\rm a\gamma\gamma}=10^{-10} \rm GeV^{-1}$, and parameters $\alpha=1.6,\;
\epsilon^4=0.15,\;
N_{\rm Xe}=2.8\times10^{29},\; R_{\rm Xe}=2.16\mathring{A}$, $N_{\rm N,0/PA}=2.42\times10^{39},\; R_{\rm N}=1.55\mathring{A},\;N_{\rm N,0/AN}\simeq 1.3\times10^{41},\;h=37\;\rm{km},\;N_O=3.1\times10^{37},\; R_{O}=1.52\;\mathring{A}.$}
    \label{fig:all}
\end{figure*}

We consider the proposed next-generation experiment XENON-LUX-ZEPLIN-DARWIN (XLZD) as an illustrative example of the sensitivity of direct detection experiments to ARPs.
XLZD will use Liquid Xenon as detector material \cite{XLZD_2025}. Liquid Xenon has a higher density than LAr, higher atomic $Z$ and mass $A$ numbers, and a larger radius and mass. XLZD is planned to have a liquid xenon target mass of 60 tonnes at the initial operational stage, which will subsequently be upgraded to 80 tonnes. This is equivalent to $2.8 \times 10^{29}$ Xe atoms initially, which is about 8 orders of magnitude less than the number of oxygen atoms in IceCube. The calculation of the event rate is analogous to what we did for IceCube. In particular, we can apply Eq.~\eqref{eq:event-rate2}, replacing the oxygen atom with the xenon atom as the target. This amounts to using the parameter values: 
$$(N_O, R_{\rm atom})  \to (N_{\rm Xe}=2.8 \times 10^{29}, R_{\rm atom} = 1.5 \mathring{A}) \; .$$
Applying the form factor formula presented in App.~\ref{ap:form-factors},\footnote{We stress again that the form factor obtained via the relativistic Hartree-Fock approximation and displayed in App.~\ref{ap:form-factors} might not capture the full structure of atomic dynamics, however it serves as a good estimate for the event rate predicted by the ARPs at these axion energies.} and Eq.~\eqref{eq:event-rate2} for the above numbers, we find that the event rate for (XLZD) is suppressed compared to IceCube for the parameter range of interest, as shown in Fig.~\ref{fig:all}. Furthermore, the ARP signal, triggered by the injection of a single high-energy photon, is highly dissimilar to the prompt scintillation signals (S1) and delayed ionisation signals (S2) that direct detection experiments are constructed to detect. It is therefore likely that an ARP signal at a direct detection experiment would be interpreted as background and discarded. However, even if the ARP signal was searched for in direct detection experiments, we do not expect it to compete with IceCube's sensitivity, given the latter's low background, large volume and long exposure.

%%%%%%%%%%%%%%%%%%%%%%%%%%%%%%%%%%%%%%%%%%%%%%%%%%%%

\section{Conclusions}
In this work, we 
have determined how axions trapped in spherical pocket regions interact with electromagnetic fields. In contrast to the much-studied case of free axions, trapped axions do not admit asymptotic states, and the standard S-matrix calculations are inapplicable. Nevertheless, we have shown that the axion-photon conversion rate in constant electromagnetic fields can be calculated analytically by expressing the axion field using spherical harmonics. Our main results are \eqref{eq:power} and the simple high-energy limit \eqref{eq:highk}, which respectively give the power or rate of emitted photons from an ARP in a constant electric field. 

We then applied our formalism to subatomic dark-matter ARPs passing through the Coulomb fields of atoms (modelled using the relativistic Hartree-Fock approximation) in terrestrial detector volumes. Axion-photon conversion leads to the emission of high-energy photons that promptly trigger electromagnetic cascades in the TeV -- EeV range. The signal is isotropic in the lab frame, and exotic upward-going electromagnetic cascades provide a smoking-gun signal of the model.  

Using these results, we surveyed classes of terrestrial detectors that may be sensitive to ARPs. We showed that recent data from IceCube constrains a small part of the parameter space. Balloon-borne radio observatories, such as ANITA and PUEO, as well as the Pierre Auger Observatory, are only sensitive to the smallest pockets (where the axions have the highest energies), for which the event-rate is strongly suppressed. Furthermore, traditional direct detection experiments are less sensitive because of their smaller detector volume.  

Our work suggests several natural extensions. First, in this paper, we only consider the simplest models of ARP dark matter in which the axions are sufficiently weakly coupled to never co-thermalise with the Standard Model. More strongly coupled axions may yield stronger signals, but require a careful reanalysis of their cosmic formation and present-day properties, as we outlined in Sec.~\ref{sec:thermalised}. Second, we have only considered terrestrial experiments in this paper. Large-scale astrophysical environments with strong magnetic fields will also generate ARP signals and may be used to search for ARPs. Third, we have only considered axions coupled to photons, but additional couplings to e.g.~electrons are possible and may lead to interesting phenomenology. We expect to return to these questions in future work.   

In sum, ARPs provide a novel, viable dark matter theory with a phenomenology that is distinct from existing paradigms. Progress in developing the theory further can be made through theoretical modelling, phenomenological applications and direct confrontation with observational data. We hope that the results presented in this paper may provide the tools and grounding for such investigations.   

\acknowledgments
We thank Itay Bloch, Aleksandr Chatrchyan, Ariel Goobar, Takeshi Kobayashi, Thong Nguyen, Anna Obertacke and Philip S{\o}rensen for illuminating discussions. DM and CN are supported by the Swedish Research Council (VR) under grants 2019-02337 and 2024-04289. This article is based upon work from COST Action COSMIC WISPers CA21106, supported by COST (European Cooperation in Science and Technology). WK would like to acknowledge support from the ICTP through the associates Programme (2022-2027). CN presented this work in the ``Planck2026 and 6th EuCAPT Symposium'', the COST Action COSMIC WISPers 4th training school, the ``Invisibles 2026'' workshop and the ``Cosmo2026'' conference and thanks the participants for interesting discussions.

\appendix
\section{Overlap Integral}
\label{app:overlap-integral}
We observe that our calculation boils down to the calculation of the overlap integral $\mathbf{F}^{nlm}_{\mathbf{\Omega}}$ within the spherical pocket, which in spherical coordinates is
\begin{equation}\mathbf{F}^{nlm}_{\mathbf{\Omega}}=g_{a\gamma\gamma}\int d^3r e^{-i\mathbf{k}_\gamma\cdot \mathbf{r}}\mathbf{E}\times \nabla u_{nlm}(\mathbf{r})\;.
\end{equation}
For the following, we will assume that the electric field is approximately constant inside the pocket $\mathbf{E}(\mathbf{x})\simeq\mathbf{E}_0$. The integral can be solved to higher orders in a perturbative expansion of $\mathbf{E}(\mathbf{x})$, using properties of the Legendre polynomials in the multipole expansion. We can integrate by parts, use the Dirichlet boundary conditions to obtain
\begin{align}
    &\mathbf{F}^{nlm}_{\mathbf{\Omega}}=\int d^3 r e^{-i\mathbf{k}_\gamma\cdot \mathbf{r}}\mathbf{E}_{0}\times \nabla u_{nlm}=\\&=-\int d^3 re^{-i\mathbf{k}_\gamma\cdot \mathbf{r}}\nabla\times\left(u_{nlm}\mathbf{E}_0\right) = \\&=-\int d^3 r\;\nabla \times\left(e^{-i\mathbf{k}_\gamma\cdot \mathbf{r}}u_{nlm}\mathbf{E}_0\right)-i\int d^3 r e^{-i\mathbf{k}_\gamma\cdot \mathbf{r}}u_{nlm}\mathbf{k}_\gamma\times \mathbf{E}_{0}\; \\&=-\oint_{\partial V}d\mathbf{S}\times \left(\mathbf{E}_0\;e^{-i\mathbf{k}_\gamma\cdot \mathbf{R}} \; u_{nlm}(\mathbf{R})\right)-i \int d^3 r\;\mathbf{k}_\gamma \times  \mathbf{E}_0e^{-i\mathbf{k}_\gamma\cdot \mathbf{r}} u_{nlm}(\mathbf{r})\;.
\end{align}

From the boundary conditions we have $u_{nlm}(\mathbf{R)=}0$. Then

\begin{align}
\mathbf{F}^{nlm}_{\mathbf{\Omega}}=-i \mathbf{k}_\gamma \times\mathbf{E}_{0}\int d^3 r \;e^{-i\mathbf{k}_\gamma\cdot \mathbf{r}} \;u_{nlm}(\mathbf{r)}\;,
\end{align}
In the case of a Coulomb-like field (e.g a radial field inside atoms), we would have the equation $\mathbf{E}_{\rm atom}=-\nabla \Phi_{\rm atom}$. The fact that $\mathbf{E}_{\rm atom}$ is curl-free, allows us to arrive at the same conclusion, without assuming that the field is constant inside the pocket
\begin{align}
\mathbf{F}^{nlm}_{\mathbf{\Omega}}=-i \mathbf{k}_\gamma \times\int d^3 r \;e^{-i\mathbf{k}_\gamma\cdot \mathbf{r}} \mathbf{E}_{\rm atom}(\mathbf{d+r}) \;u_{nlm}(\mathbf{r)}\;,
\end{align}
with $\mathbf{d}$ the distance from the atomic centre. For the ARP sizes being considered, we have that for the most part of the parameter space $|\mathbf{d}|\gg R_{\rm pocket}$, such that $\mathbf{E}_{\rm atom}(\mathbf{d}+\mathbf{r})\simeq \mathbf{E}_{\rm atom}(\mathbf{d})$ and then 
\begin{align}
\mathbf{F}^{nlm}_{\mathbf{\Omega}}=i\mathbf{E}_{\rm atom}(\mathbf{d})\times \mathbf{k}_\gamma \int d^3 r \;e^{-i\mathbf{k}_\gamma\cdot \mathbf{r}} \;u_{nlm}(\mathbf{r})\;.
\end{align}

We can expand the plane wave 
\begin{equation}
    e^{-i\mathbf{k}_\gamma\cdot \mathbf{r}}=4\pi \sum_{l'=0}^{\infty}\sum_{m'=-l'}^{m'=l'}i^{l'} Y_{l'm'}(\hat{\mathbf{k}}_\gamma) Y^*_{l'm'}(\mathbf{\hat{r}}) j_{l'}(k_\gamma r)\;,
\end{equation}
and using properties of spherical harmonics, we reduce the volume integral to a radial integral
\begin{align}
&\mathbf{F}^{nlm}_{\mathbf{\Omega}}=4\pi i^{l+1}\mathcal{C}_{nl}\;   \;\mathbf{E}_{\rm atom}(\mathbf{d})\times \mathbf{k}_\gamma\; Y_{lm}(\mathbf{\hat{k}_\gamma})\int_0^{R_{\rm pocket}} dr\;r^2 \; j_l(k_\gamma r)j_l(k_{nl} r)\;.
\end{align}

The latter integral is analytically known, and so we arrive at a closed form expression for $\mathbf{F_{\Omega}}$
\begin{equation}
\mathbf{F}^{nlm}_{\mathbf{\Omega}}=4\pi i^{l+1}\mathcal{C}_{nl}\;   \mathbf{k}_\gamma\times \mathbf{E}_{\rm atom}(\mathbf{d})\; Y_{lm}(\mathbf{\hat{k}_\gamma}) \frac{R_{\rm pocket}^2 k_{nl}}{k_{nl}^2-k_\gamma^2}j_{l-1}\left(k_{nl}R_{\rm pocket}\right)j_{l}\left(k_\gamma R_{\rm pocket}\right)\;.
\end{equation}
\subsection{High-Energy Limit}
The discussion will start from the axion-photon conversion rate, calculated using the quantum field theory of axions inside a sphere
\begin{equation}
    \Gamma_{a\rightarrow \gamma}^{nlm}=\frac{g_{a\gamma\gamma}^2\omega^2}{16\pi^2 }\mathcal{B}_{nl}^2E^2_{0}\int d\Omega_k \sin^2\psi \Big|\int d^3r e^{-i\mathbf{k_\gamma} \mathbf{r}} u_{nlm}(\mathbf r)\Big|^2\;.
\end{equation}
Furthermore, we assume we have prepared a fixed $(n,l)$ state, thus the amplitude has a degeneracy on m, such that $\mathcal{M}^2\rightarrow \frac{1}{2l+1}\mathcal{M}^2$, resulting to
\begin{equation}
\Gamma_{a\rightarrow \gamma}^{nl}=\frac{g_{a\gamma\gamma}^2\omega^2}{16\pi^2 \left(2l+1\right)}\mathcal{B}_{nl}^2E^2_{0}\int d\Omega_k \sin^2\psi \Big|\int d^3r e^{-i\mathbf{k_\gamma} \mathbf{r}} \sum_mu_{nlm}(\mathbf r)\Big|^2\;.
\end{equation}
 We now focus on the overlap integral 
\begin{align}
&\notag\sum_m\int d^3r e^{-i\mathbf{k_\gamma} \mathbf{r}} u_{nlm}(\mathbf r)=\\&=4\pi \sum_{l'm'm} i^l \int d\Omega_{\hat{\mathbf{r}}} Y_{l'm'}^*(\hat{\mathbf{r}}) Y_{lm}(\hat{\mathbf{r}})\int dr r^2 j_l(k_a r) j_{l'}(k_\gamma r) Y_{l'm'}(\hat{\mathbf{k}}_\gamma)\\&=4\pi i^l \sum_m Y_{lm}(\hat{\mathbf{k}}_\gamma)\frac{-R_{\rm pocket}^2}{k_\gamma^2-k_a^2}\left[k_a j_l\left(k_\gamma R_{\rm pocket}\right)j_{l-1}(k_a R_{\rm pocket})\right]\\&\simeq 4\pi i^l \sum_m Y_{lm}(\hat{\mathbf{k}}_\gamma)\frac{-R_{\rm pocket}^2}{(k_\gamma-k_a)2k_a}k_a \frac{\sin(k_\gamma R_{\rm pocket}-l\pi/2)}{k_\gamma R}\frac{\cos(k_a R_{\rm pocket}-l\pi/2)}{k_a R}\\&\simeq 4\pi i^l \sum_m Y_{lm}(\hat{\mathbf{k}}_\gamma)\frac{-\sin(qR_{\rm pocket})}{2q k_a k_\gamma}\;,
\end{align}
where we used the fact that at that limit, the roots of the Bessel function satisfy $j_l(k_aR)=0\simeq \sin(k_a R-l\pi/2)$. Note that we suppressed the indices $n,l$ on $k_a$. 

Squaring and performing the 3j integral as above, the result reduces to 
\begin{equation}
\Gamma_{a\rightarrow \gamma}^{nl}=\frac{g_{a\gamma\gamma}^2\mathcal{B}_{nl}^2}{6}E^2_{0}\frac{\sin^2(qR_{\rm pocket})}{q^2 k_a^2}=\frac{g_{a\gamma\gamma}^2 E^2_{0}}{3 R}\frac{\sin^2(qR_{\rm pocket})}{q^2}\;.
\end{equation}
We used that $B_{nl}^2=2k_a^2/R$ in the $k_a R\gg1$ limit. The momentum transfer is $q=\omega-\sqrt{\omega^2-m_a^2}\simeq m_a^2/(2\omega)$, and taking the limit $m_a/\omega\rightarrow 0$, the result is independent of energy and equal to
\begin{equation}
 \Gamma_{a\rightarrow \gamma}^{nl}=\frac{g_{a\gamma\gamma}^2 E^2_{0} R_{\rm pocket}}{3}\;.
\end{equation}
\section{Detailed event rate derivations}
\label{ap:event-rate}
In this appendix, we provide detailed calculations related to event rates for different detectors. A general simplification for all detectors is the following integral
\begin{equation}
    \int_{b_{\rm min}}^{R_{\rm atom}} db \;b\int_{t_{\rm in}}^{t_{\rm out}} dt E^2_{\rm atom}\;.
\end{equation}
Changing variables from t to $r=\sqrt{b^2+v_{\rm DM}^2t^2}$, we end up with the following double integral
\begin{equation}
   \int_{b_{\rm min}}^{R_{\rm atom}} db \;b\int_{t_{\rm in}}^{t_{\rm out}} dt E^2_{\rm atom}= \frac{2}{v_{\rm DM}}\int_{b_{\rm min}}^{R_{\rm atom}} db \;b\int_{b}^{R}dr \frac{r}{\sqrt{r^2-b^2}} E^2_{\rm atom}\;.
\end{equation}
We observe that we scan a triangle defined by $b_{\rm min}\leq b\leq R,\; b\leq r\leq R$, and thus we can swap the integration variables by 
\begin{equation*}
   \frac{2}{v_{\rm DM}}\int_{b_{\rm min}}^{R_{\rm atom}} db \;b\int_{b}^{R}dr \frac{r}{\sqrt{r^2-b^2}} E^2_{\rm atom}= \frac{2}{v_{\rm DM}}\int_{b_{\rm min}}^{R_{\rm atom}} dr \;\int_{b_{\rm min}}^{r}db \frac{rb}{\sqrt{r^2-b^2}} E^2_{\rm atom}(r)\;.
\end{equation*}
Integrating over b, we get
\begin{equation}
    \int_{b_{\rm min}}^{R_{\rm atom}} db \;b\int_{t_{\rm in}}^{t_{\rm out}} dt E^2_{\rm atom}= \frac{2}{v_{\rm DM}}\int_{b_{\rm min}}^R dr\; r \sqrt{r^2-b_{\rm min}^2} E^2_{\rm atom}(r)\;.
\end{equation}
We will use this form for all following calculations.
\subsection{IceCube}
In the following part we will approximate that the main target atoms in the detector are the oxygen atoms, so $n_{\rm atom}$ and $E_{\rm atom}$ refer to oxygen. Furthermore, we will use the high-energy limit for the emission rate $\Gamma_{a\rightarrow\gamma}$ calculated in Eq.~\ref{eq:highk}. We will also use a geometric cross-section of the form $\sigma_{\rm target}=\pi\left( R_{\rm atom}^2-b_{\rm min}^2\right)$ for the atomic cross-section for the atom-ARP interaction.  The event rate is then
\begin{align}
    &\nonumber\dot{N}_{\rm events}=\frac{\pi }{6}g_{a\gamma\gamma}^2n_{\rm pocket}v_{\rm DM}n_{\rm atom}A_{\rm det}\langle L\rangle R_{\rm pocket}N_a\\&\times\int_{b_{\rm min}}^{R_{\rm atom}} db \;b\int_{t_{\rm in}}^{t_{\rm out}} dt E^2_{\rm atom}\;,
\end{align}

For IceCube (and any convex volume), we can use the mean-chord theorem to write $\langle L\rangle =4V_{\rm det}/A_{\rm det}$. Furthermore, the number density of ARPs is given by $n_{\rm pocket}=1/D^3$, with D the pocket mean separation. This results in
\begin{align}
    &\nonumber\dot{N}_{\rm events}=\frac{2\pi  }{3D^3 }g_{a\gamma\gamma}^2v_{\rm DM}n_{\rm atom}V_{\rm det}\\&\times R_{\rm pocket}N_a\int_{b_{\rm min}}^{R_{\rm atom}} db \;b\int_{t_{\rm in}}^{t_{\rm out}} dt E^2_{\rm atom}\;,
\end{align}
or simply since $N_{\rm atom}=n_{\rm atom} V_{\rm det}$ 
\begin{equation}
   \dot{N}_{\rm events}=\frac{2\pi }{3D^3 }g_{a\gamma\gamma}^2v_{\rm DM}N_{\rm atom} R_{\rm pocket}N_a\int_{b_{\rm min}}^{R_{\rm atom}} db \;b\int_{t_{\rm in}}^{t_{\rm out}} dt E^2_{\rm atom}\;.
\end{equation}
Swapping the integrals and using 
\begin{equation}
 E_{\rm atom}^2=\alpha_{\rm EM}/4\pi \left(\frac{Z_{\rm atom}-N_e(r)}{r^2}\right)^2\;,
\end{equation}
we arrive at
\begin{align}
&\dot{N}_{\rm events}=\frac{\alpha_{\rm EM}  }{3D^3 }g_{a\gamma\gamma}^2\\&\times N_{\rm atom} R_{\rm pocket}N_a\int_{b_{\rm min}}^{R_{\rm atom}} dr \;r\sqrt{r^2-b_{\rm min}^2} \left(\frac{Z_{\rm atom}-N_e(r)}{r^2}\right)^2\;.
\end{align}

\subsection{ANITA}
\label{ap:anita}
The payload flies at height $h$ above the ice. We use a spherical coordinate system and denote the zenith angle by $\theta$, chosen so that the payload is located at $\theta=0$ and $r=R_E+h$. The rate is then given by
\bea
 \dot{N} &= 
  \frac{1}{4}  n_{\rm pocket}  \langle N_{\gamma,\rm pass}\rangle v_{\rm pocket}\sigma_{i\rm atom} n^{(0)}_{i,\rm atm} \DOz
   \int_{R_E}^{R_E+h} dr \int_0^{\theta_{\rm max}(r)} d\theta \int_0^{2\pi} d\phi r^2 \sin \theta \;  e^{-2\left(\frac{r-R_E}{H_s}\right)} \nonumber \\
&=   
 \frac{\pi}{2}  n_{\rm pocket} \langle N_{\gamma,\rm pass}\rangle v_{\rm pocket}\sigma_{i\rm atom} n^{(0)}_{i,\rm atm} \DOz
   \int_{0}^{h} dz\; (R_E+z)^2  e^{-2\left(\frac{z}{H_s}\right)}\left(
   1- \cos(\theta_{\rm max}(z)) \right)
   \nonumber \\
&=   
\nonumber \frac{\pi}{2}  n_{\rm pocket} \langle N_{\gamma,\rm pass}\rangle v_{\rm pocket}\sigma_{i\rm atom} n^{(0)}_{i,\rm atm} \DOz R_E^3
   \int_{0}^{\tilde h} d \tilde z\; (1+ \tilde z)^2  e^{-2\left(\frac{\tilde z}{\tilde H_s}\right)}\left(
   1- \cos(\theta_{\rm max}(z)) \right)\, , 
\eea
where $\tilde z= z/R_E$, $\tilde h= h/R_E$, and $\tilde H_s= H_s/R_E$. 
The maximal zenith angle of the spherical surface at height $\tilde z$ within the detector's sub-horizon field of view is given by
$$
\cos\theta_{\rm max}(z) = \frac{(1+ \tilde h)^2+ (1 + \tilde z)^2- \left( \sqrt{\tilde h^2 + 2 \tilde h} - \sqrt{\tilde z^2 + 2 \tilde z} \right)^2}{2(1+ \tilde h)(1+\tilde z)} \, .
$$
Since $\tilde h,\, \tilde H_s \ll 1$, we approximate the integrand to linear order in $\tilde z,\, \tilde h,\, \tilde H_s$, so that
\bea
\dot{N}&= \frac{\pi}{4}  n_{\rm pocket} \langle N_{\gamma,\rm pass}\rangle v_{\rm pocket}\sigma_{i\rm atom} n^{(0)}_{i,\rm atm} \DOz R_E^3
   \int_{0}^{\tilde h} d \tilde z\;\left( \sqrt{2\tilde h} - \sqrt{2 \tilde z}\right)^2  e^{-2\left(\frac{\tilde z}{\tilde H_s}\right)}\, \nonumber \\\notag
   &= \frac{\pi}{8}  n_{\rm pocket} \langle N_{\gamma,\rm pass}\rangle v_{\rm pocket}\sigma_{i\rm atom} n^{(0)}_{i,\rm atm} \DOz R_E^2 H_s \\\notag & \times
   \left( 
   2 \tilde h + \tilde H_s\left(1-{\rm exp}(-2 h/H_s)\right) - \sqrt{2\pi \tilde h \tilde H_s} {\rm Erf}(\sqrt{2 h/H_s})\nonumber  \right) \\
   \approx& 
   \frac{\pi}{8}  n_{\rm pocket} \langle N_{\gamma,\rm pass}\rangle v_{\rm pocket}\sigma_{i\rm atom} n^{(0)}_{i,\rm atm} \DOz R_E H_s
   \left( 
   2  h +  H_s - \sqrt{2\pi \tilde h \tilde H} \;.
   \right)
\eea
Plugging the value for $\langle N_\gamma\rangle_{\rm pass}$, we can approximate the event rate as
\begin{align}
\notag&&\dot{N}(h) =\frac{\alpha\pi g_{\rm a\gamma\gamma}^2}{12}\frac{N_aR_{\rm pocket}}{D^3}n_{N,0}\Delta\Omega^{0}R_E H_s\left(2h+H_s-\sqrt{2\pi\tilde{h}\tilde{H_s}}\right)\\&&\times\int_{b_{\rm min}}^{R_{\rm atom}} dr \;r\sqrt{r^2-b_{\rm min}^2} \left(\frac{Z-N_e(r)}{r^2}\right)^2\simeq\\&&\notag\simeq \frac{\alpha\pi g_{\rm a\gamma\gamma}^2}{12}\frac{N_aR_{\rm pocket}}{D^3}n_{N,0}\Delta\Omega^{0}R_E H_s\left(2h+H_s\right)\\&&\times\int_{b_{\rm min}}^{R_{\rm atom}} dr \;r\sqrt{r^2-b_{\rm min}^2} \left(\frac{Z-N_e(r)}{r^2}\right)^2 \;.
\end{align}
\subsection{Pierre Auger}
\label{ap:PA}
The idealised rate for Pierre Auger is given by 
\begin{equation}
   \dot{N}=\int dV\int d\Omega \frac{d^2n_\gamma}{dtd\Omega}\;.
\end{equation}
Using the expressions for the distribution of photons and the effective axion-photon conversion rate, we see that
\begin{equation*}
   \dot{N} =\frac{8\pi g_{\rm a\gamma\gamma}^2}{3} N_an_{\rm pocket}R_{\rm pocket} \int dV n_N(\mathbf{x}) \int_{b_{\rm min}}^{R_{\rm atom}} dr \;r\sqrt{r^2-b_{\rm min}^2} E^2_{\rm atom}(r)\;.
\end{equation*}
The volume integral can be reduced to
\begin{align}
 &\notag \dot{N}=\frac{2\alpha g_{\rm a\gamma\gamma}^2}{3}\frac{N_aR_{\rm pocket}}{D^3}S_{\rm det}\\&\times \int_0^{H_s} dz n_N^0 \exp\left(-\frac{z}{H_s}\right) \int_{b_{\rm min}}^{R_{\rm atom}} dr \;r\sqrt{r^2-b_{\rm min}^2} \left(\frac{Z_{\rm atom}-N_e(r)}{r^2}\right)^2
\end{align}
and performing the integral gives
\begin{align}
 &&\notag \dot{N} =\frac{2\alpha g_{\rm a\gamma\gamma}^2}{3}\frac {R_{\rm atom}^2}{R_{\rm atom}^2-b_{\rm min}^2} \frac{N_aR_{\rm pocket}}{D^3} n_{N,0}V_{\rm atm}(1-e^{-H/H_s})\\&&\times\int_{b_{\rm min}}^{R_{\rm atom}} dr \;r\sqrt{r^2-b_{\rm min}^2} \left(\frac{Z_{\rm atom}-N_e(r)}{r^2}\right)^2\;.
\end{align}
\section{Review of ``Quantum Mechanical'' calculation for plane waves}
\label{ap:plane-waves}
We will review the calculation of axion-photon conversion in the Wentzel-Kramers-Brillouin (WKB) approximation assuming that the geometry of the system is such that allows plane wave solutions for the axion \cite{PhysRevD.37.1237,Marsh:2021ajy}. We will perform the calculation for a constant background electric field, which we denote as $\mathbf{E}_{0}$.  

We linearise the fields as 
\begin{align}
   &a= a(\mathbf{x,t})\;,\\
   &\mathbf{E}=\mathbf{E}_{0}+\mathbf{e}\;,\\
   &\mathbf{B}=\mathbf{b}\;.
\end{align}
The modified Maxwell equations are, assuming quantities depend only on the direction of axion propagation -z- and that both axion and photon fields have energy $\omega$,
\begin{align}
    &\left(\Box+m_a^2 \right)a=-g_{a
    \gamma\gamma}\epsilon_{ijk}E_{0,i}\partial_j A_k=g_{a
    \gamma\gamma}\left(E_{0,x}\partial_zA_y-E_{0,y}\partial_zA_x\right)\,\\
     &\left(\Box+m_{\rm pl}^2 \right)A_x=g_{a
    \gamma\gamma}E_{0,y}\partial_z a\;,\\
    &\left(\Box+m_{\rm pl}^2 \right)A_y=-g_{a
    \gamma\gamma}E_{0,x}\partial_z a\;.
\end{align}
The equations can be written as
\begin{align}
    &\left(\omega^2+\partial_z^2-m_a^2 \right)a=-g_{a
    \gamma\gamma}\left(E_{0,x}\partial_zA_y-E_{0,y}\partial_zA_x\right)\,\\
     &\left(\omega^2+\partial_z^2-m_{\rm pl}^2 \right)A_x=-g_{a
    \gamma\gamma}E_{0,y}\partial_z a\;,\\
    &\left(\omega^2+\partial_z^2-m_{\rm pl}^2 \right)A_y=g_{a
    \gamma\gamma}E_{0,x}\partial_z a\;,
\end{align}
and we use the approximation $\omega^2+\partial_z^2=2\omega\left(\omega-i\partial_z\right)$. We furthermore write the dispersion relation for the axion as $k_a=n\omega=\omega+\left(n-1\right)\omega$, with $|n-1|\ll 1$, such that we can write $g_{a\gamma\gamma}\partial_za\simeq ig_{a\gamma\gamma}\omega a$ and then after shifting the axion $a\rightarrow ia$, we get
\begin{align}
    &\left(\omega-i\partial_z\right)a-\frac{m_a^2}{2\omega}a=-\frac{g_{a\gamma\gamma}}{2}\left(E_xA_y-E_yA_x\right),\;\\
    &\left(\omega-i\partial_z\right)A_x-\frac{m_{\rm pl}^2}{2\omega}A_x=\frac{g_{a\gamma\gamma}}{2}E_ya,\\
    &\left(\omega-i\partial_z\right)A_y-\frac{m_{\rm pl}^2}{2\omega}A_y=-\frac{g_{a\gamma\gamma}}{2}E_xa,\;
\end{align}
with $E_i\equiv E_{0,i}$. This equation can be written as
\begin{equation}
    i\partial_z\Psi=\left(H_0+H_I\right)\Psi,\;
\end{equation}
with
\begin{equation}
    H_I=\begin{pmatrix}
0 & 0 & -\Delta_y \\
0 & 0& \Delta_x \\
-\Delta_y & \Delta_x & 0
\end{pmatrix},\; \Psi=\begin{pmatrix}
A_x \\ A_y \\ a
\end{pmatrix},\; H_0=\omega\mathbf{1}+\begin{pmatrix}
\Delta_\gamma & 0 & 0 \\
0 & \Delta_\gamma& 0 \\
0 & 0 & \Delta_a
\end{pmatrix},
\end{equation}
and $\Delta_i=\frac{g_{\rm a\gamma\gamma}}{2}E_i$,\; $\Delta_{a/\gamma}=-m_{\rm a/pl}^2/2\omega$. This is equivalent to the Schrödinger equation and thus we can use similar perturbative techniques to solve it (see \cite{PhysRevD.37.1237,Marsh:2021ajy} for more details). 

For the case of constant $\mathbf{E}=\mathbf{E}_0$, we can solve it in a straightforward way by diagonalising the Hamiltonian. Starting from a pure axion state $\Psi(0)=(0,0,1)^T$, we get
\begin{equation}
    \mathcal{A}_{a\rightarrow \gamma_x}=(1,0,0)\Psi(L),\;
\end{equation}
and thus
\begin{equation}
   P_{a\rightarrow\gamma}=\left( |\mathcal{A}_{a\rightarrow \gamma_x}|^2+ |\mathcal{A}_{a\rightarrow \gamma_y}|^2\right)=\frac{4\Delta_{\perp}^2}{4\Delta_\perp^2+\tilde\Delta^2}\sin^2\left(\frac{\lambda_3-\lambda_2}{2}L\right)
\end{equation}
with $\Delta_\perp^2=\Delta_x^2+\Delta_y^2$, $\tilde\Delta=\Delta_\gamma-\Delta_a$ and the difference between eigenvalues is
\begin{equation}
    \lambda_3-\lambda_2=\sqrt{\tilde{\Delta}^2+\Delta_{\perp}^2}\;.
\end{equation}
So, the probability of conversion for an axion travelling at a distance L
\begin{equation}
    P_{a\rightarrow\gamma}=\frac{\Theta^2}{1+\Theta^2}\sin^2\left(\Delta\sqrt{1+\Theta^2}L\right)
\end{equation}
with $\Theta=\frac{2g_{a\gamma\gamma}E_{\perp}\omega}{m_{\rm eff}^2},\; \Delta=
    \frac{m_{\rm eff^2}}{4\omega}\;$.
\paragraph{Perturbation Theory}
In the case that the electric field is approximately constant at the region of interest, i.e
\begin{equation}
    \mathbf{E}\left(\mathbf{x}\right)=\mathbf{E}(\mathbf{x_0})+\left(\mathbf{x}-\mathbf{x}_0\right)\cdot\nabla\mathbf{E}|_{\mathbf{x}_0}+\dots\equiv \mathbf{E}_0+\dots\;,
\end{equation}
we can use perturbation theory to solve the "Schrödinger" equation
\begin{equation}
    i\partial_z\Psi=\left(H_0+H_I\right)\Psi,\;
\end{equation}
with $H_I$ being z- dependent, involving the full electric field $\mathbf{E}(\mathbf{x})$. In order to do this, we go the interaction picture and use perturbation theory, so 
\begin{equation}
    \Psi_{\rm int}(z)=\Psi_{\rm int}(0)-i\int_0^z dz' H_I \Psi_{\rm int}(z)
\end{equation}
so we have to calculate
\begin{align}
    \mathcal{A}_{a\rightarrow\gamma_x}=(1,0,0)\Psi(L)=i\int_0^L dz' e^{-i\left(\Delta_a-\Delta_\gamma\right)z'}\Delta_y(z')\;,\\
     \mathcal{A}_{a\rightarrow\gamma_y}=(0,1,0)\Psi(L)=-i\int_0^L dz' e^{-i\left(\Delta_a-\Delta_\gamma\right)z'}\Delta_x(z')\;.
\end{align}
At leading order in perturbation theory, the $\Delta_{a,\gamma}$ are independent of z, so the result can be obtained trivially
\begin{equation}
    |\mathcal{A}_{a\rightarrow\gamma}|^2=g_{a\gamma\gamma}^2 E_\perp^2\frac{\sin^2\left[\tilde \Delta L/2\right]}{\tilde \Delta ^2}
\end{equation}
\section{Data for Electric Fields}
\label{ap:form-factors}
We use the tabulated data for the atomic form factor of oxygen (via the fitting function also used in \cite{Abe_2021}) found in \cite{Brown2004}. The form factors were acquired using the relativistic Hartree-Fock approximation.
\begin{equation}
    F_O(q)=\sum_{i=1}^4a_i \exp\left[-b_i\left(\frac{|\mathbf{q}|}{4\pi}\right)^2\right]+c,\;
\end{equation}
Tabulated data on the form factor as well as the parameters $(a_i,b_i,c)$ are given \cite{Brown2004}.  Note that the fitting formula is valid for $0<q<25 \mathring{A}^{-1}$.

For oxygen, nitrogen and xenon the fitting parameters are given by the following table for the low-q regime

\begin{center}
\begin{tabular}{ |c||c||c| } 
\hline
\multicolumn{3}{|c|}{Oxygen} \\
\hline
 $a_1=3.04850$ & $b_1=13.2771\; \mathring{A}^{2}$ & $c=0.25080$ \\ 
 $a_2=2.28680 $ & $b_2=5.70110\; \mathring{A}^{2}$ & \\ 
 $a_3=1.54630$ & $b_3=0.32390\; \mathring{A}^{2}$ & \\ 
 $a_4=0.86700$ & $b_4=32.9089\; \mathring{A}^{2}$ & \\
\hline \multicolumn{3}{|c|}{Nitrogen} \\
 \hline
 $a_1=12.2126$ & $b_1=0.005700\; \mathring{A}^{2}$ & $c=-11.529$ \\ 
 $a_2=3.13220$ & $b_2=9.89330\; \mathring{A}^{2}$ & \\ 
 $a_3=2.01250$ & $b_3=28.9975\; \mathring{A}^{2}$&\\ 
 $a_4=1.16630$ & $b_4=0.582600\; \mathring{A}^{2}$& \\
 \hline \multicolumn{3}{|c|}{Xenon} \\
 \hline
 $a_1=20.2933 $ & $b_1=3.92820\; \mathring{A}^{2}$ & $c=3.71180$ \\ 
 $a_2=19.0298$ & $b_2=0.34400\; \mathring{A}^{2}$ & \\ 
 $a_3=8.97670$ & $b_3=26.4659\; \mathring{A}^{2} $&\\ 
 $a_4=1.99000$ & $b_4=64.2658\; \mathring{A}^{2}$& \\
 \hline
\end{tabular}
\end{center}

\section{Constant Magnetic fields}
\label{ap:magn}
Here the interaction term $\mathbf{j}_a$ is
\begin{equation}
    \mathbf{j}_a=-g_{\rm a\gamma\gamma}\mathbf{B}_0\partial_t a=i\omega_{a,nl}
   g_{\rm a\gamma\gamma}\mathbf{B}_0 a\;,
\end{equation}
so that
\begin{equation}
     \frac{dP}{d\Omega}=\langle \mathbf{\hat{n}}\cdot \left(\mathbf{E}\times\mathbf{B}\right)\rangle r^2=\frac{ k_{\gamma}\omega_{a,nl}}{8\pi^2}|\hat{\mathbf{n}}\times \mathbf{F_\Omega}|^2\;.
\end{equation}
and
\begin{equation}
    \mathbf{F_\Omega}=g_{a\gamma\gamma} \mathbf{B}_0 i\omega_{a,nl}\int d^3 r e^{-i\mathbf{k}_\gamma \mathbf{r}}u_{{nlm}}(\mathbf{r})
\end{equation}
As before
\begin{equation}
|\mathbf{\hat{n}}\times \mathbf{B}_0|^2=\mathbf{B}_0^2\sin^2\psi\;,
\end{equation}
we have already calculated the spherical integral of the mode function. Thus, since $k_\gamma=\omega_{a,nl}$, the result is identical to the electric field case with $\mathbf{E}_{\rm atom}\rightarrow \mathbf{B}_0$.

\bibliographystyle{JHEP}
\bibliography{bibl}
\end{document}